\documentclass[times,5p,twocolumn]{elsarticle} 
\usepackage[numbers]{natbib}

\usepackage{framed}

\usepackage{amssymb}
\usepackage{latexsym}
\usepackage{subfiles}
\usepackage{makecell}
\usepackage{lineno}

\usepackage{amsmath,booktabs,siunitx}
\usepackage{enumitem}
\usepackage{algorithm}
\usepackage[noend]{algpseudocode}
\algrenewcommand\textproc{\FuncCall}
\usepackage{mathtools}

\newcommand{\norm}[1]{\left\lVert#1\right\rVert}
\newcommand{\FuncCall}[1]{\texttt{#1}}
\newcommand{\hrulealg}[0]{\vspace{1mm} \hrule \vspace{1mm}}

\usepackage{url}
\usepackage{xcolor}
\usepackage{pifont}
\newcommand{\cmark}{\color{green}\ding{51}}%
\newcommand{\xmark}{\color{red}\ding{55}}%

\usepackage{hyperref}
\definecolor{newcolor}{rgb}{.8,.349,.1}

\journal{Medical Image Analysis}

\begin{document}


\begin{frontmatter}

\title{Robust automated calcification meshing for biomechanical cardiac digital twins}



\author[1]{Daniel H. Pak\corref{cor1}}
\cortext[cor1]{Corresponding author}
\ead{daniel.pak@yale.edu}
\author[2]{Minliang Liu}
\author[1]{Theodore Kim}
\author[3]{Caglar Ozturk}
\author[4]{Raymond McKay}
\author[3]{Ellen T. Roche}
\author[2]{Rudolph Gleason}
\author[1]{James S. Duncan}

\address[1]{Yale University, 300 Cedar St, New Haven, CT 06511, USA}
\address[2]{Texas Tech University, 805 Boston Avenue, Lubbock, TX, 79409}
\address[3]{Massachusetts Institute of Technology, 45 Carleton St, Cambridge, MA 02142 USA}
\address[4]{Hartford Hospital, 85 Seymour St, Hartford, CT 06106 USA}
\address[5]{Georgia Institute of Technology, 315 Ferst Dr NW, Atlanta, GA 30332 USA}



\begin{abstract}
Calcification has significant influence over cardiovascular diseases and interventions. Detailed characterization of calcification is thus desired for predictive modeling, but calcified heart meshes for physics-driven simulations are still often reconstructed using manual operations. This poses a major bottleneck for large-scale adoption of computational simulations for research or clinical use. To address this, we propose an end-to-end automated meshing algorithm that enables robust incorporation of patient-specific calcification onto a given heart mesh. The algorithm provides a substantial speed-up from several hours of manual meshing to $\sim$1 minute of automated computation, and it solves an important problem that cannot be addressed with recent template registration-based heart meshing techniques. We validated our final calcified heart meshes with extensive simulations, demonstrating our ability to accurately model patient-specific aortic stenosis and Transcatheter Aortic Valve Replacement. Our method may serve as an important tool for accelerating the development and usage of physics-driven simulations for cardiac digital twins. 









\end{abstract}

\end{frontmatter}


Calcification plays an important role in cardiovascular diseases (CVD). High levels of calcification in the coronaries, aorta, and heart valves have all been shown to be effective predictors of CVD incidence and death \citep{greenland2004coronary,chen2017coronary,witteman1986aortic,nicoll2014predictive}, and vascular calcification has been correlated with atherosclerosis  \citep{sangiorgi1998arterial,durham2018role}. For aortic stenosis (AS), degenerative calcification and the subsequent over-stiffening of the aortic valve leaflets is the most common etiology in industrialized countries \citep{mohler2004mechanisms}, and calcification is a diagnostic biomarker for inconclusive AS \citep{pawade2019and}.

Beyond simple quantification and statistical analyses, physics-driven biomechanical simulations can help elucidate the effects of calcification on cardiovascular physiology. The morphometry of calcification can significantly affect the patient-specific hemodynamics across the aortic valve \citep{ge2010direction,halevi2016fluid}, and the irregular stress patterns induced by calcification may stimulate further progression of calcification \citep{weinberg2010hemodynamic,arzani2017strain,qin2020role}. Calcification can also heavily influence the outcome of Transcatheter Aortic Valve Replacements (TAVR), potentially causing various complications, such as aortic rupture, paravalvular leak, and conduction abnormalities \citep{milhorini2020valvular,pollari2020aortic}. Simulation-based cardiac digital twins have been actively explored as a tool for assessing these risks and planning for TAVR procedures \citep{wang2015simulations,sturla2016impact}.

Accurate modeling of calcification has thus been a long-standing clinical interest, leading to notable progress in automated algorithms for calcification segmentation. For non-contrast computed tomography (CT), global thresholding and filtering via aorta segmentation has been suggested for calcification segmentation \citep{kurugol2015automated}. For CT angiography (CTA), where the lumen intensity is much higher and may vary across different patients, adaptive thresholding based on luminal attenuation is the most common approach \citep{mahabadi2009association,alqahtani2017quantifying,bettinger2017practical,vlastra2019aortic}. Learning-based methods have also been proposed using hand-crafted features \citep{grbic2013image,harbaoui2016aorta} and deep learning \citep{graffy2019automated}.

The caveat is that incorporating calcification into simulations often requires converting the predicted segmentation into a finite element mesh. The conversion generally consists of extensive manual work, primarily due to the complexity of the final mesh and the unreliability of automated meshing operations \citep{wang2015simulations,grbic2013image}. Manual adjustments could take several hours per model for a trained expert, which introduces a key bottleneck in the simulation workflow. Alternatives exist, such as assigning different element properties on the heart tissue geometry \citep{morganti2014simulation,loureiro2020biomechanical}, defining surface-based tie constraints \cite{russ2013simulation}, and re-meshing calcification with a nearby tissue \citep{bianchi2019patient}. However, existing workarounds are suboptimal due to various reasons, such as missing geometry, instability in complex simulations, and inability to setup large-scale simulations due to the loss of the heart mesh correspondence.

In contrast, we propose a robust fully automated calcification meshing algorithm that satisfies the following challenging constraints:

\begin{itemize}[noitemsep]
    \item Accurate calcification geometry
    \item Clean manifold surface for tetrahedralization
    \item Preservation of the input heart mesh topology
    \item Complete surface matching (i.e. node-to-node and edge-to-edge) along the heart-calcification contact surfaces
\end{itemize}

A key benefit of our approach is that it enables large-scale physics-driven simulations with complex multi-body interactions. The initial geometry definition as well as the simulation setup are both massively simplified due to our output mesh characteristics, resulting in a speed-up of up to a few orders of magnitude for simulation setup. Together with the recent advances in heart tissue meshing algorithms, our method could serve as an important tool for accelerating the development and usage of physics-driven simulations for cardiac digital twins.



 \label{sec:01_intro}

\section*{Results}
\subsection*{Heart mesh reconstruction}

For automated meshing of the calcified heart, we must consider two main components with substantially different geometrical characteristics: the native heart tissue and the calcification. For the native heart tissue, we modeled the aortic valve, ascending aorta, and left ventricular myocardium. This allowed us to test our algorithm on a sizable portion of the ascending aorta with the complex aortic valve geometry included in the domain.

From the numerous recently proposed algorithms \citep{kong2021deep,kong2022learning,pak2021distortion,pak2021weakly}, we decided to use a slightly modified version of DeepCarve for heart mesh reconstruction \citep{pak2023patient}, mainly to take advantage of its demonstrated performance on aortic valve meshing.  Here, we briefly describe the method for completeness.

The main objective for DeepCarve is to train a deep learning model that will take an input image and generate a patient-specific mesh by deforming a template mesh. The required training label is a set of component-specific pointclouds that can help measure the spatial accuracy of the prediction. The predicted mesh is penalized for large deformations from the original template elements via deformation gradient-based isotropic and anisotropic energies. The deformation generated by the model is constrained to be a diffeomorphic b-spline field to enforce additional smoothness.

The three minor modifications for this work are (1) random translation augmentation, (2) a better LV template with fewer distorted elements, and (3) the addition of 0.1*AMIPS \citep{fu2015computing} to the training loss. These changes mainly allowed us to adapt to more realistic meshing environments, while maintaining similar model performance.

Once DeepCarve is trained, a single forward-pass of the 3D patient scan directly generates high-quality volumetric meshes of the heart geometry in $<$1 second. We confirmed the accuracy of the reconstructed heart geometry (Fig.~\ref{fig:combined_overview}a, Fig.~\ref{fig:combined_seg_results}a) and focused on robustly incorporating calcification to the resulting output.  For additional detail, we refer the readers to the original publication \citep{pak2023patient}.

\subsection*{Calcification Meshing with Anatomical Consistency}

\begin{figure*}[!t]
    \centerline{\includegraphics[width=\linewidth]{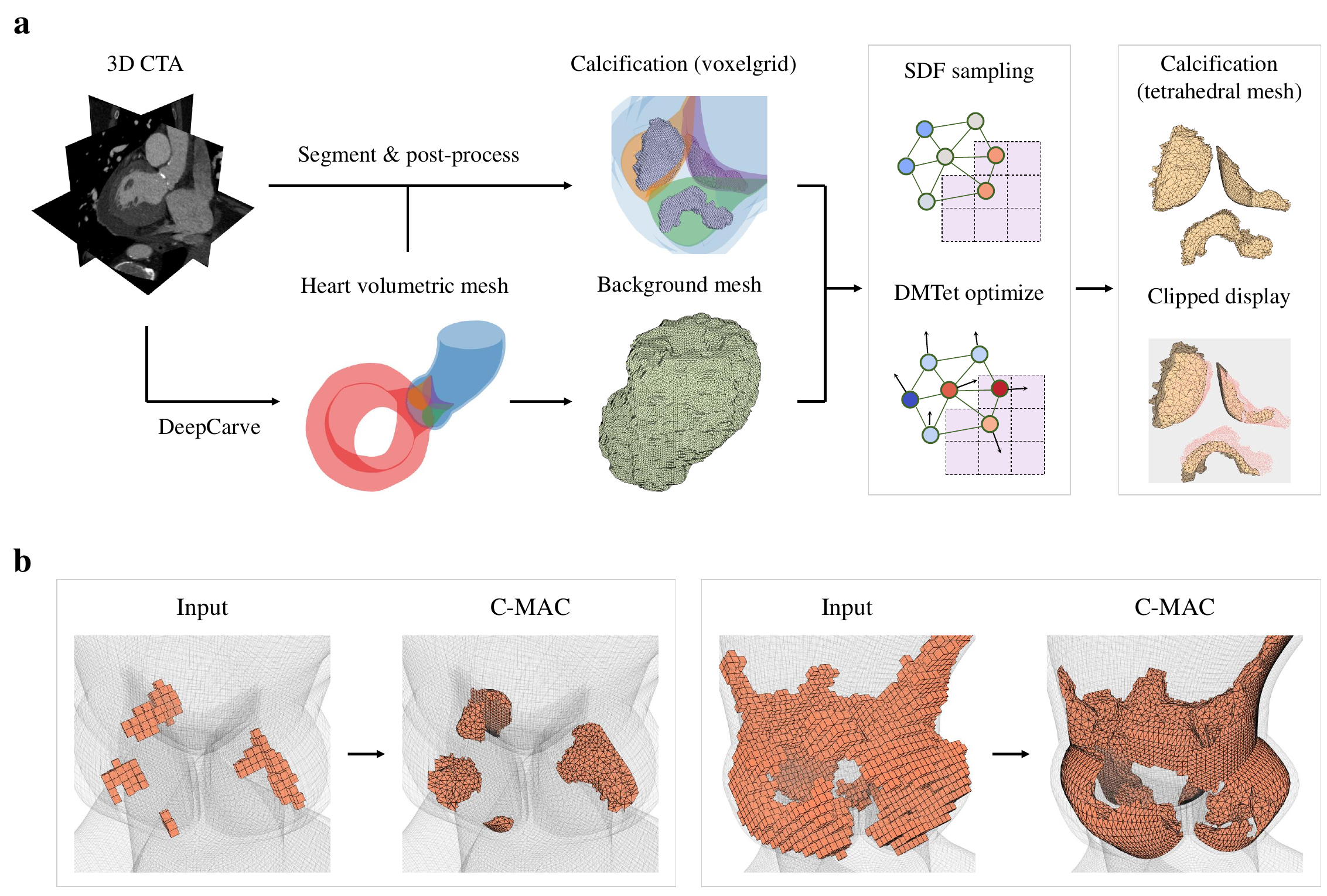}}
    \caption{(a) The main overview of C-MAC. Starting from a 3D CTA, a patient-specific heart volumetric mesh is first generated using DeepCarve. The resulting heart mesh is used to aid the voxelgrid segmentation of calcification, as well as to generate the background mesh for DMTet. The SDF values are first initialized by sampling the voxelgrid segmentation at each node of the background grid, and subsequently the node coordinates and nodal SDF values are optimized for better element quality. The resulting DMTet output mesh is further processed via node-based remeshing to generate the final input for tetrahedralization. The details of each process can be found in the main text and later figures. (b) C-MAC robustly and automatically incorporates a voxelgrid segmentation into an existing mesh without changing the original mesh topology. Here, we demonstrate its performance using two vastly different calcification segmentations (orange) and complex aortic valve meshes (gray).}
    \label{fig:combined_overview}
\end{figure*}

Our calcification meshing algorithm requires two inputs: the original patient CTA scan and the reconstructed heart mesh. The main goal is to incorporate calcification into the heart mesh for more accurate and realistic structural simulations. The major steps are outlined in Fig.~\ref{fig:combined_overview}a and Algorithm~\ref{alg:c-mac}, and samples of input/output pairs are shown in Fig.~\ref{fig:combined_overview}b.

A deep learning model is used to perform the initial calcification segmentation, and a series of model-based post-processing techniques are used to ensure the anatomical consistency between the predicted segmentation and the reconstructed heart mesh. The segmentation and the heart mesh are then processed intricately together to generate the final calcification mesh. The meshing steps are centered around Deep Marching Tetrahedra and controlled mesh simplification via constrained remeshing. The details and the performance of each step will be discussed in the following sections. We will henceforth refer to our algorithm as C-MAC (\textbf{C}alcification \textbf{M}eshing with \textbf{A}natomical \textbf{C}onsistency).

\begin{algorithm}
    \caption{C-MAC overview} \label{alg:c-mac}
    \begin{algorithmic}[1]
    \Function{C-MAC}{$I, M_{heart}$}
    \State $y_{0} \gets \FuncCall{DLCa2Seg}(I)$
    \State $y_{ca2} \gets \FuncCall{PostProcessCa2Seg}(y_0, M_{heart})$
    \State $M_{bg} \gets \FuncCall{GenerateBackgroundMesh}(M_{heart})$ 
    \State $S_{ca2} \gets \FuncCall{DMTetOpt}(y_{ca2}, M_{bg})$
    \If{not satisfied}
    \State $S_{ca2} \gets \FuncCall{ConstrainedRemesh}(S_{ca2})$
    \EndIf
    \State $M_{ca2} \gets \FuncCall{TetGen}(S_{ca2})$
    \State \Return{$M_{ca2}$}
    \EndFunction
    \end{algorithmic}
    \hrulealg
    $I$: image, $y$: segmentation, $M$: volume mesh, $S$: surface mesh
\end{algorithm}

\subsection*{Initial segmentation via deep learning \textnormal{(\FuncCall{DLCa2Seg})}}

\begin{figure*}[!t]
    \centerline{\includegraphics[width=\linewidth]{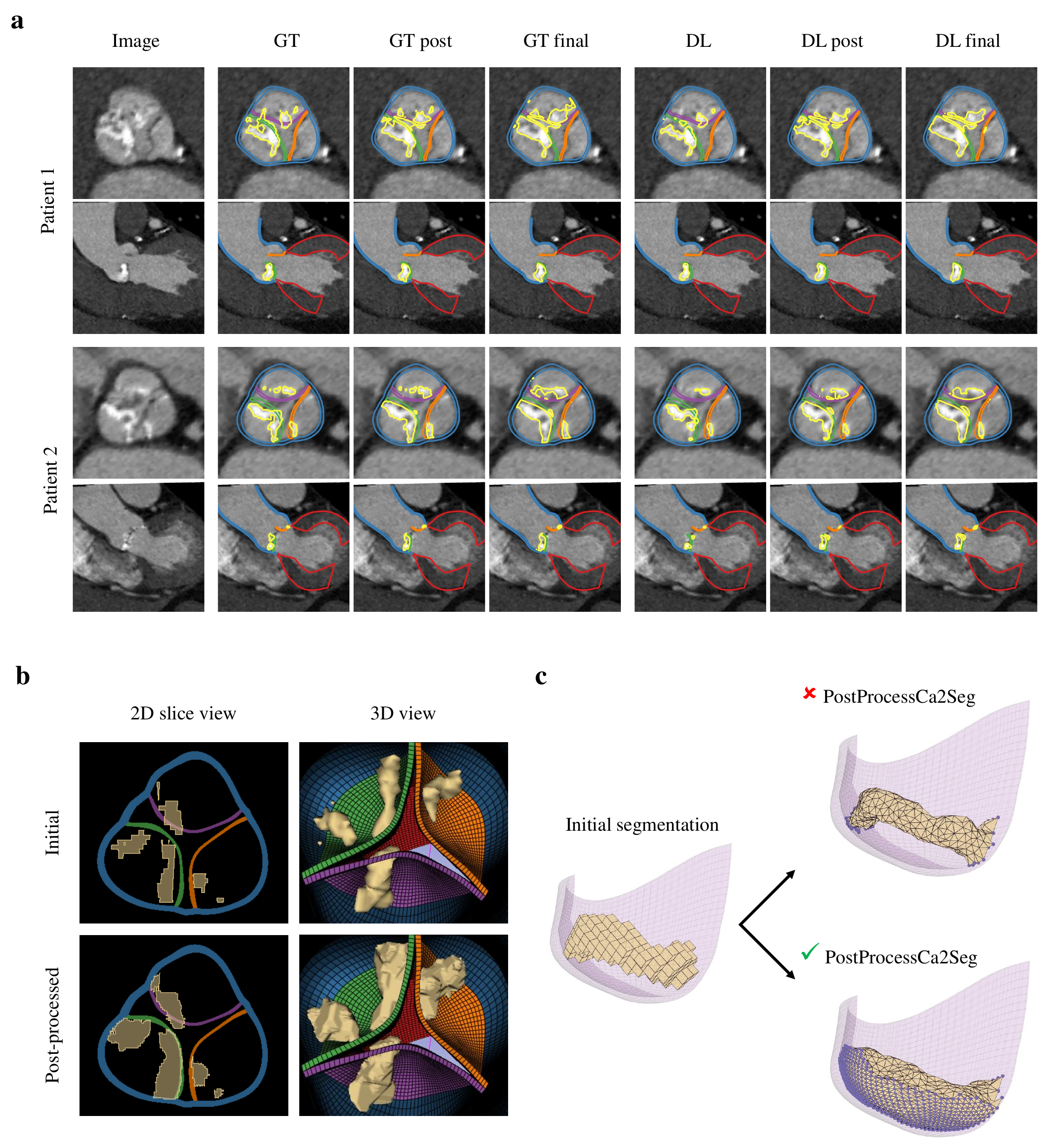}}
    \caption{(a) Qualitative evaluation of the spatial accuracy of calcification using two image slice views for two test-set patients. GT: ground-truth segmentation, DL: deep learning segmentation using ``GDL (ours)", post: result of \FuncCall{PostProcessCa2Seg} using either GT or DL as the initial segmentation, final: result of the full \FuncCall{C-MAC}. Yellow: calcification, red: partial LV myocardium, blue: aorta, (green, orange, purple): aortic valve leaflets. (b) Highlighting the effects of \FuncCall{PostProcessCa2Seg} on one test-set patient. The first two columns are two different 2D-slice views, and the third column is a 3D view of the aortic valve. (c) Visualizing the effect of \FuncCall{PostProcessCa2Seg} on the final C-MAC mesh. Purple dots indicate merged nodes between the calcification and the heart mesh. This example illustrates both the benefit and drawback of our post-processing algorithm. Benefit: improved anatomical consistency with the surrounding heart tissue. Drawback: some overestimation of calcified regions.}
    \label{fig:combined_seg_results}
\end{figure*}

Given the inherent voxelgrid structure of 3D medical images, the most common intermediate representation for meshing is voxelgrid segmentation. Following this trend, we first focused on accurately extracting the initial segmentation of the calcified regions within the CTA.

Thresholding techniques are easy to implement and generally perform well for calcification segmentation. However, they are particularly vulnerable to imaging artifacts due to their simplistic approach, and they are usually supplemented with manual post-processing to refine the segmentation. Deep learning (DL) models, on the other hand, utilize learned feature aggregation to make the final prediction, freeing themselves from the simple errors of thresholding approaches. Therefore, for our task, we aimed to train a DL model that can handle the following conditions:

\begin{itemize}[noitemsep]
    \item Highly unbalanced target (i.e. calcification is very sparse compared to the input image volume)
    \item Empty target (i.e. no calcification in the input image)
    \item Accurate segmentation boundary
\end{itemize}

To meet these criteria, we used a modified 3D U-net \citep{ronneberger2015u} and evaluated four different training strategies involving the following losses:

\begin{itemize}[noitemsep]
    \item Dice + Cross Entropy (DiceCE)
    \item Generalized Dice Loss (GDL)
    \item GDL + boundary loss (Gbd)
    \item GDL + chamfer loss (Gch)
\end{itemize}



The evaluations are summarized in Table~\ref{table:results_dl_seg} (top). We first observed that nnU-net \cite{isensee2021nnu}, a popular self-configurable segmentation algorithm, suffers from a high rate of false positive segments outside of the aorta. All of our methods had better overall performance than nnU-net. The only metric that nnU-net performed relatively well was the number of false positives for empty targets (i.e. when there is no aortic calcification in the CTA scan).

Among our own models, we first compared the effects of different image intensity ranges for the min-max normalization, while using DiceCE or GDL. The metrics generally indicate that (1) using the right intensity range can help boost the performance and (2) GDL with $I$ minmax [-200,1500] achieves the best overall performance.

Then, we used the best performing training setup from our initial evaluation and tested the modified GDL losses: Gbd and Gch. Gbd was effective at reducing false positives, but otherwise both Gbd and Gch showed worse overall performance than the standard GDL loss. For our task, false positives for empty targets were less critical than accurate prediction of non-empty targets because false positives for empty targets can be easily identified and omitted using the source image. Therefore, we used the GDL with $I$ minmax [-200,1500] as our final segmentation model for downstream tasks.

Fig.~\ref{fig:combined_seg_results}a shows a few representative examples of ground-truth segmentation vs. DL segmentation using ``GDL (ours)" (column 2 vs. column 5). The DL segmentation qualitatively matches the ground-truth very well.

\begin{table*}[t!]
\caption{Deep learning model performances on the initial calcification segmentation (\FuncCall{DLCa2Seg}) with and without post-processing (\FuncCall{PostProcessCa2Seg}). $I$ minmax: range for image intensity min-max normalization, Post: post-processing, HD: Hausdorff distance, CD: mean symmetric chamfer distance, CD$_{\textrm{Heart}}$: one-sided chamfer distance from calcification to heart surfaces, False pos: false positives, mm: millimeters, vx: voxels, $*$: $p<0.05$ between GDL (ours) and nnU-net. All metrics are mean $\pm$ stdev across patients.}
\centering
\label{table:results_dl_seg}

\def\sym#1{\ifmmode^{#1}\else\(^{#1}\)\fi}
\sisetup{detect-weight,mode=text}
\robustify\bfseries

\begin{tabular*}{\linewidth}{ l @{\extracolsep{\fill}} c c *{4}{S[table-format=1.3(3),separate-uncertainty=true,table-align-text-post=false]} S[table-format=2.1(4),separate-uncertainty=true,table-figures-uncertainty=1]}

\toprule

\multicolumn{3}{c}{Experimental condition} & \multicolumn{4}{c}{Non-empty target segmentation} & \multicolumn{1}{c}{Empty target} \\
\cmidrule(lr){1-3} \cmidrule(lr){4-7} \cmidrule(lr){8-8}


{\makecell{Train strategy}} & {\makecell{$I$ minmax}} & {\makecell{Post}} & {\makecell{Dice $\uparrow$}} & {\makecell{HD 95\% (mm) $\downarrow$}} & {\makecell{CD (mm) $\downarrow$}} & {\makecell{CD$_{\textrm{Heart}}$ (mm) $\downarrow$}} & {\makecell{False pos (vx) $\downarrow$}}\\

\midrule

nnU-net & self-configured & \xmark &           0.656 \pm 0.222    &           13.259 \pm 18.055    &           2.337 \pm 2.994    &           2.501 \pm 2.955    &             5.4 \pm  10.0 \\
\midrule[0pt]
DiceCE & [ -158,  864]   & \xmark &          0.685 \pm 0.180    &            8.254 \pm 11.605    &           1.462 \pm 2.421    &           1.345 \pm 0.319    &            25.5 \pm  25.2 \\
DiceCE & [ -200, 1500]   & \xmark &          0.690 \pm 0.144    &            8.362 \pm 13.141    &           1.320 \pm 1.653    &           1.653 \pm 0.922    &            28.9 \pm  30.1 \\
DiceCE & [-1000, 2500]   & \xmark &          0.680 \pm 0.134    &            5.684 \pm  8.165    &           1.155 \pm 1.529    &           1.409 \pm 0.301    &           151.9 \pm 249.2 \\
\midrule[0pt]
GDL     & [ -158,  864]   & \xmark &          0.685 \pm 0.184    &            6.503 \pm  9.066    &           1.380 \pm 2.102    &           1.381 \pm 0.419    &            48.9 \pm  81.0 \\
GDL (ours)& [ -200, 1500]   & \xmark & \bfseries 0.709 \pm 0.111    & \bfseries  4.668 \pm  6.977$^{*}$ & \bfseries 0.842 \pm 1.081$^*$  & \bfseries 1.314 \pm 0.280$^*$  &            31.5 \pm  44.2 \\
GDL     & [-1000, 2500]   & \xmark &          0.679 \pm 0.152    &            7.336 \pm  9.745    &           1.419 \pm 1.932    &           1.711 \pm 1.320    &           129.2 \pm 171.6 \\
\midrule[0pt]
Gbd     & [ -200, 1500]   & \xmark &          0.675 \pm 0.142    &            5.797 \pm  7.180    &           1.000 \pm 1.002    &           1.319 \pm 0.295    & \bfseries   4.4 \pm   4.8 \\
Gch     & [ -200, 1500]   &  \xmark &         0.672 \pm 0.150    &            6.306 \pm  8.282    &           1.098 \pm 1.089    &           1.365 \pm 0.306    &            16.2 \pm  14.3 \\

\midrule 

nnU-net    & self-configured & \cmark &          0.629 \pm 0.158 &           8.315 \pm 10.688 &           1.478 \pm 1.910 & \bfseries 0.929 \pm 0.188 & \bfseries  2.2 \pm  6.4 \\
DiceCE     & [-200, 1500]    & \cmark &          0.657 \pm 0.086 &           5.058 \pm  7.658 &           0.918 \pm 0.860 &           0.989 \pm 0.219 &           24.9 \pm 41.7 \\
GDL (ours) & [-200, 1500]    & \cmark & \bfseries 0.657 \pm 0.079 &           5.083 \pm  7.277 & \bfseries 0.834 \pm 0.758 &           0.976 \pm 0.206 &           22.8 \pm 31.0 \\
Gbd        & [-200, 1500]    & \cmark &          0.632 \pm 0.113 &           4.973 \pm  5.074 &           0.906 \pm 0.644 &           0.979 \pm 0.209 &            2.9 \pm  4.1 \\
Gch        & [-200, 1500]    & \cmark &          0.624 \pm 0.112 & \bfseries 4.543 \pm  4.421 &           0.900 \pm 0.611 &           0.972 \pm 0.200 &            5.2 \pm  5.4 \\

\bottomrule

\end{tabular*}


\end{table*}


\subsection*{Segmentation post-processing \textnormal{(\FuncCall{PostProcessCa2Seg})}} \label{sec:results_seg_post}

Although our DL model delivered promising initial results, it still falls short of the high standards required for simulations. In particular, the predicted segmentation may have slight spatial gaps (2-3 voxels) to the input heart mesh, which are generally negligible errors but not for establishing contacts between the predicted calcification and the surrounding heart tissue. These small errors could drastically alter the results of the simulations, so we aimed to explicitly improve the anatomical consistency of the DL predictions by removing any obvious gaps between the calcification segmentation and the input heart mesh.




Our main method for filling in the gaps is morphological closing (dilation $\rightarrow$ erosion). Unfortunately, a na\"ive application of this approach with differently sized isotropic filters leads to overly connected components, such as undesired calcification fusing and excessive contact with the neighboring heart structures. We incorporated several additional operations, such as the combination of isotropic and anisotropic filters and voxelgrid upsampling, to minimize the deviation from the initial segmentation while still being able to fill in the gaps.

For segmentation post-processing, we first evaluated its effects on the final segmentation's spatial accuracy and anatomical consistency (Table~\ref{table:results_dl_seg} bottom). The overall improvements in the mean surface accuracy metric (CD) combined with our qualitative evaluations (Fig.~\ref{fig:combined_seg_results}a) indicate that our algorithm reasonably preserves the initial segmentation's spatial characteristics. Significant improvements in the calcification-to-heart distance (CD$_{\textrm{heart}}$) indicates reduced spatial gaps and better anatomical consistency between calcification and its surrounding tissue.

The improved surface metrics HD and CD demonstrate our algorithm's effectiveness in filtering out easily avoidable false-positive calcification segments, such as those outside the aorta and those with tiny volumes. Interestingly, this filtering process reduced the performance gap between different segmentation models (Table.~\ref{table:results_dl_seg} top $\rightarrow$ Table.~\ref{table:results_dl_seg} bottom), which suggests that our post-processing algorithm was effective at removing segmentation outliers for various models.


The worse Dice and better CD$_{\textrm{Heart}}$ demonstrates the trade-off between volume overlap and anatomical consistency. This is expected, since the current strategy for removing the calcification-to-heart spatial gaps is to extend the calcification segments while keeping the heart mesh fixed. For the purposes of calcification meshing, the benefit of anatomical consistency far outweighs the slight reduction in volume overlap (Fig.~\ref{fig:combined_seg_results}b, Fig.~\ref{fig:combined_seg_results}c). An interesting future direction is to jointly optimize both the heart mesh and the calcification segmentation to encourage anatomical consistency while minimizing this trade-off.

\subsection*{Background mesh generation \textnormal{(\FuncCall{GenerateBackgroundMesh})}}

The post-processed segmentation and the input heart mesh are combined together to perform calcification meshing, and Deep Marching Tetrahedra (DMTet) \citep{shen2021deep} is at the core of our meshing algorithm. Similar to regular marching tetrahedra, DMTet generates an isosurface triangular mesh given a background tetrahedral mesh and its nodal SDF. Unlike regular marching tetrahedra, DMTet additionally performs simultaneous optimization of the background mesh node positions and the nodal SDF to refine the final mesh.

DMTet defines the typology of the isosurface based on the sign changes of the SDF and the location of the isosurface based on the magnitude of the SDF (Eq. \ref{eq:dmtet_v_ab}).

\begin{equation} \label{eq:dmtet_v_ab}
    v_{ab}^{\prime} = \dfrac{v_a \cdot SDF(v_b) - v_b \cdot SDF(v_a)}{SDF(v_b) - SDF(v_a)}
\end{equation}

\noindent Here, $v_i$ is the nodal position and $SDF(v_i)$ is the nodal SDF. Note that the node interpolation is only performed along edges where there is a ``sign" change, i.e. between a positive SDF node and a non-positive SDF node ($SDF(v_a) \leq 0$ and $SDF(v_b) > 0$). This prevents division by 0.


Based on the DMTet definition, the background mesh is a main driver of the final output mesh topology. To enable contact surface matching, we designed our background mesh to be constrained by the original heart surfaces, with a few minor adjustments to enable closed non-overlapping calcification surfaces (Fig.~\ref{fig:combined_mesh_results}a). We first extracted the relevant heart surfaces from the input heart mesh, and then generated a clean manifold offset layer from the extract surfaces to add to our input to TetGen \citep{hang2015tetgen}. Following tetrahedralization, the elements inside the original heart surface were removed, and fake tetrahedral elements were added to the boundary surfaces to ensure closed calcification surfaces at the end of the DMTet tessellation.

The success rate for background mesh tetrahedralization was 100\%, mainly due to the nice heart mesh quality from DeepCarve and our method of generating a clean manifold offset surface. The density of the tetrahedral elements was tuned to provide enough resolution, but the element quality was not critical, as we additionally refined the surfaces in the following steps.

\subsection*{DMTet optimization \textnormal{(\FuncCall{DMTetOpt})}}

We evaluated the remaining meshing steps of C-MAC using the following metrics: (1) the success rate of tetrahedralization, (2) the spatial accuracy of the final mesh surface, and (3) the element quality of the final tetrahedra (Table.~\ref{table:meshing_techniques}). We also computed the same metrics for existing mesh editing approaches to showcase our improvements. For all methods, the general strategy was to first construct a triangular surface mesh using the post-processed segmentation and the heart mesh, and then apply TetGen to tetrahedralize the inner volume.

For baselines, we applied mesh boolean operations from a wide variety of existing meshing libraries. For 5 of the 8 libraries, the success rate for the final tetrahedralization was below 10\% (Table.~\ref{table:meshing_techniques}). For some, even the initial boolean operation failed completely. This is consistent with the notion that mesh boolean operations are generally unreliable, especially when coupled with complex input geometry and complex downstream tasks such as constrained tetrahedralization.

\begin{figure*}[!t]
    \centerline{\includegraphics[width=\linewidth]{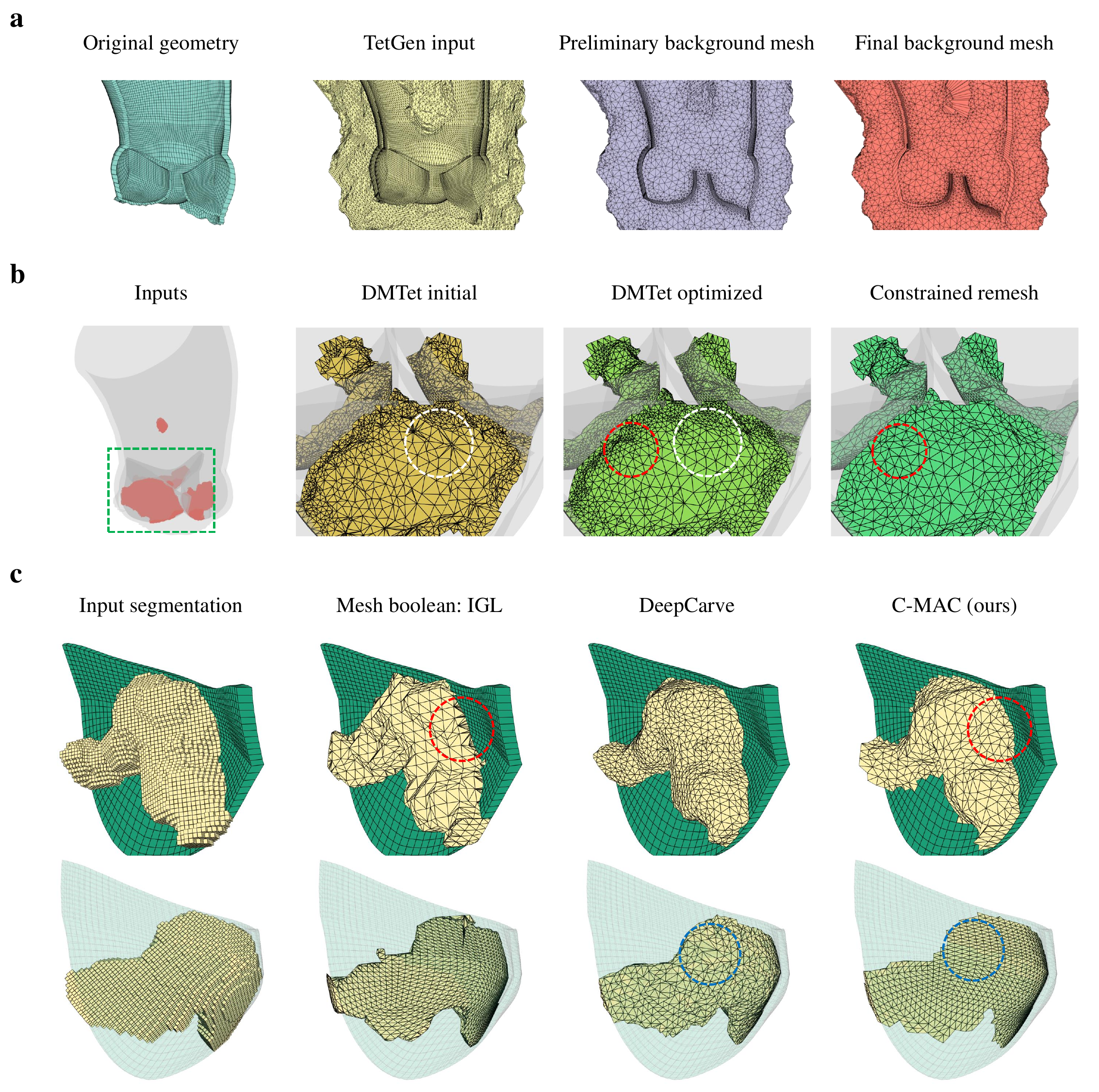}}
    \caption{(a) Illustration of the background mesh generation process. From left to right: patient-specific mesh of the aorta + aortic valve leaflets,  TetGen input generated by combining the exterior surface and an offset surface from the original geometry, preliminary background mesh generated by TetGen and hollowing, and final background mesh after adding a “fake” vertex to the exterior surface elements of the preliminary mesh. All meshes are clipped at a viewing plane for visualization purposes. (b) The three main sequential steps for anatomically consistent surface meshing. From left to right: initial inputs of patient-specific aorta + aortic valve leaflets (gray) and voxelgrid segmentation of calcification (red), initial DMTet mesh with raw sampled SDF, optimized DMTet mesh, and final remeshed surface. Green box indicates the viewing region, and colored circles indicate noticeable regions of improvement after each step. (c) Baseline comparisons for the final mesh quality. Yellow: calcification, green: aortic valve leaflet. Top: front view, bottom: back view. Colored circles indicate the noticeable regions of improvement from each baseline to C-MAC.}
    \label{fig:combined_mesh_results}
\end{figure*}

For mesh boolean operations with relatively high success rates, the issue still remained that the element qualities were heavily compromised along the mesh intersections. This is shown both qualitatively as thin triangular elements (Fig.~\ref{fig:combined_mesh_results}c), and quantitatively as low Jacobian determinants (Table.~\ref{table:meshing_techniques}). Mesh boolean approaches had worse performance in essentially every category compared to our method.


As an additional baseline, we performed DeepCarve's node stitching operations as outlined in \cite{pak2023patient}. The only metric that performed well for this method was the Jacobian determinant, which makes sense given the generally smooth and uniform elements (Fig.~\ref{fig:combined_mesh_results}c). However, the method notably suffered from irregular node-stitching along the contact surfaces, which is further confirmed by the low numbers of merged nodes between the calcification and heart meshes (Table.~\ref{table:meshing_techniques}). The irregular node-stitching also causes various other issues, such as the occasional undesired mesh intersections and failure of TetGen due to the weirdly shaped elements. Since it is difficult to automatically and robustly fix ill-formed meshes, our approach aims to generate good quality meshes from the beginning.

For marching tetrahedra, a well-known shortcoming is the irregular shape and distribution of the triangular elements along the output mesh surfaces \citep{payne1990surface,chan1998new}. This can be easily verified by applying DMTet without any optimization (Fig.~\ref{fig:combined_mesh_results}b). Such irregularity noticeably degrades the success rate of tetraheralization and the final tetrahedra quality.

Thus, we applied DMTet optimization on both the background mesh node locations and the nodal SDF values to increase the output mesh quality while minimizing the deviation from the input segmentation. In addition, we modified the nodal SDF values for two edge cases to ensure clean transition elements along the boundary and contact surfaces.

DMTet optimization provides significant improvements for all metrics (Table.~\ref{table:meshing_techniques}). DMTet optimization was crucial in reaching 100\% success rate of tetrahedralization, and it led to consistent improvements over DMTet without optimization regardless of the application of the subsequent remeshing steps. Qualitatively, it especially helps address the aggregation of sliver elements characteristic of marching tetrahedra (Fig.~\ref{fig:combined_mesh_results}b). In most cases, it was entirely acceptable to stop the meshing algorithm after DMTet optimization, given its high quality outputs.

\begin{table*}[t]
\caption{The effects of the calcification surface meshing algorithms on the final tetrahedral mesh. ``Bool" refers to the mesh boolean difference operation between the original calcification surface and the aorta + aortic valve surface. All metrics were obtained using 10 repeated trials to account for the occasional failure of TetGen. We used the post-processed ground truth segmentation as the initial segmentation. Spatial accuracy metrics were measured against the initial segmentation, so we are strictly measuring the error induced by the meshing steps. All metrics are calculated using samples with non-empty targets only. $*$: $p<0.05$ between C-MAC and ``Bool: corefinement". $\dagger$: $p<0.05$ between C-MAC and DeepCarve.}
\centering
\label{table:meshing_techniques}

\def\sym#1{\ifmmode^{#1}\else\(^{#1}\)\fi}
\sisetup{detect-weight,mode=text}
\robustify\bfseries

\begin{tabular*}{\linewidth}{ l @{\extracolsep{\fill}} S[table-format=3.1] *{3}{S[table-format=1.3(3),separate-uncertainty=true,table-align-text-post=false]} S[table-format=4(4),separate-uncertainty=true,table-align-text-post=false] S[table-format=1.3(3),separate-uncertainty=true,table-align-text-post=false]}

\toprule

{\makecell{Mesh technique}} & {\makecell{Success (\%}) $\uparrow$} & {\makecell{Dice $\uparrow$}} & {\makecell{HD 95\% (mm) $\downarrow$}} & {\makecell{CD (mm) $\downarrow$}} & {\makecell{Merged nodes} $\uparrow$} & {\makecell{$\mid J \mid$ \;$\uparrow$}}\\

\midrule

Bool: pymadcad     &  0.0 &   N/A           &   N/A           &   N/A           &    N/A        &   N/A           \\
Bool: OpenSCAD     &  0.0 &   N/A           &   N/A           &   N/A           &    N/A        &   N/A           \\
Bool: vtk          &  0.0 &   N/A           &   N/A           &   N/A           &    N/A        &   N/A           \\
Bool: cgal         &  3.7 &   N/A           &   N/A           &   N/A           &    N/A        &   N/A           \\
Bool: blender      &  7.4 & 0.777 \pm 0.113 & 1.207 \pm 0.293 & 0.351 \pm 0.206 &  441 \pm  546 & 0.349 \pm 0.099 \\
Bool: cork         & 70.0 & 0.794 \pm 0.060 & 1.399 \pm 0.080 & 0.384 \pm 0.081 & 3013 \pm 1742 & 0.458 \pm 0.043 \\
Bool: igl          & 77.8 & 0.791 \pm 0.066 & 1.394 \pm 0.090 & 0.370 \pm 0.080 & 2718 \pm 1570 & 0.449 \pm 0.044 \\
Bool: corefinement & 81.5 & 0.793 \pm 0.066 & 1.395 \pm 0.088 & 0.370 \pm 0.079 & 2748 \pm 1539 & 0.452 \pm 0.043 \\

\midrule[0pt]

DeepCarve         & 92.6 & 0.733 \pm 0.077 & 2.362 \pm 3.071 & 0.607 \pm 0.270 &  658 \pm 237 & \bfseries 0.605 \pm 0.020 $^{\dagger}$ \\

\midrule[0pt]

\FuncCall{DMTet} (no opt) &  89.6 & 0.814 \pm 0.044 & 1.030 \pm 0.075 & 0.234 \pm 0.017 & 3582 \pm 1995 & 0.547 \pm 0.013 \\

\midrule[0pt]

\makecell[l]{\FuncCall{DMTet} (no opt) $\rightarrow$ \\ \FuncCall{ConstrainedRemesh}} &  99.3 & 0.819 \pm 0.041 & 1.041 \pm 0.118 & 0.237 \pm 0.017 & \bfseries 3666 \pm 2024 & 0.582 \pm 0.008 \\

\midrule[0pt]

\FuncCall{DMTetOpt} & \bfseries 100.0 & 0.827 \pm 0.038 & \bfseries 1.019 \pm 0.073 & \bfseries 0.225 \pm 0.018 & 3583 \pm 1995 & 0.579 \pm 0.008 \\

\midrule[0pt]

\makecell[l]{\FuncCall{DMTetOpt} $\rightarrow$ \\ \FuncCall{ConstrainedRemesh} \\ (\FuncCall{C-MAC})} & \bfseries 100.0$^*$ & \bfseries 0.831 \pm 0.037$^{*\dagger}$ & 1.035 \pm 0.108$^{*\dagger}$ & 0.226 \pm 0.018$^{*\dagger}$ & 3576 \pm 1993$^{*\dagger}$ & 0.588 \pm 0.010$^{*}$ \\

\bottomrule

\end{tabular*}

\end{table*}

\subsection*{Constrained remeshing \textnormal{(\FuncCall{ConstrainedRemesh})}}

Although DMTet optimization already provides great quality outputs with robust performance, we can further improve the mesh quality and speed up the finite element simulations by additionally simplifying the output mesh with constrained remeshing.

The main remeshing operation is voronoi diagram-based vertex clustering \citep{valette2008generic}. By our way of constructing the background mesh and DMTet optimization, the contact surfaces already have complete surface matching to the heart mesh. To preserve this, we first split the output mesh into contact and non-contact surfaces, and only remesh the latter. For the non-contact surfaces, we further ensure that the remeshed surfaces can be combined with the original contact surfaces by preserving the border vertices during the clustering steps.

Quantitatively, constrained remeshing markedly improves the final output mesh for all evaluation metrics, especially in conditions with no DMTet optimization (Table.~\ref{table:meshing_techniques}). However, constrained remeshing without DMTet optimization is usually less reliable, as the input mesh is often riddled with aggregated sliver elements without DMTet optimization. This leads to the occasional failure of tetrahedralization as well as lower achievable final mesh quality. Constrained remeshing can also occasionally generate non-manifold surfaces due to the separation of contact and non-contact surfaces. We are still able to reach 100\% tetrahedralization success rate because we can easily store and select the optimally processed mesh at lower iterations of constrained remeshing by simply performing tetrahedralization on all intermediate meshes. This is possible because both constrained remeshing and calcification tetrahedralization take a fraction of a second to perform, and the outputs are also not very memory intensive.

As shown in Table.~\ref{table:meshing_techniques}, the condition that includes both $\FuncCall{DMTetOpt}$ and $\FuncCall{ConstrainedRemesh}$ results in the best overall meshing performance, both in terms of spatial accuracy and element quality. The final version of C-MAC is able to combine the best of both worlds of the baseline meshing methods, i.e. (1) establish complete contact surface matching, similar to mesh boolean operations, and (2) provide great element quality along the non-contact surfaces, similar to DeepCarve's approach. (Fig.~\ref{fig:combined_mesh_results}b, Fig.~\ref{fig:combined_mesh_results}c). In addition, C-MAC ensures that any node residing on any of the heart surface's planes will be coincident with an existing heart mesh node, which prevents any potential issues with minute mesh intersections that can introduce cumbersome errors during simulations.

\subsection*{Run-time} \label{sec:run_time}

The approximate run-time to generate the entire calcified heart geometry (i.e. DeepCarve + C-MAC) is $\sim$1 minute per 3D image. The three most time-consuming tasks are (1) closing operations for segmentation post-processing ($\sim$10 seconds), (2) TetGen for background mesh generation ($\sim$15 seconds), and (3) DMTet optimization ($\sim$20 seconds). Most other individual processes typically take $<$1 second. All times were measured on a single NVIDIA RTX3080Ti laptop workstation. Note that the algorithms are fully automated with no human annotations for any of the meshing steps.

\subsection*{Solid mechanics simulations}

\begin{figure*}[!t]
    \centerline{\includegraphics[width=\linewidth]{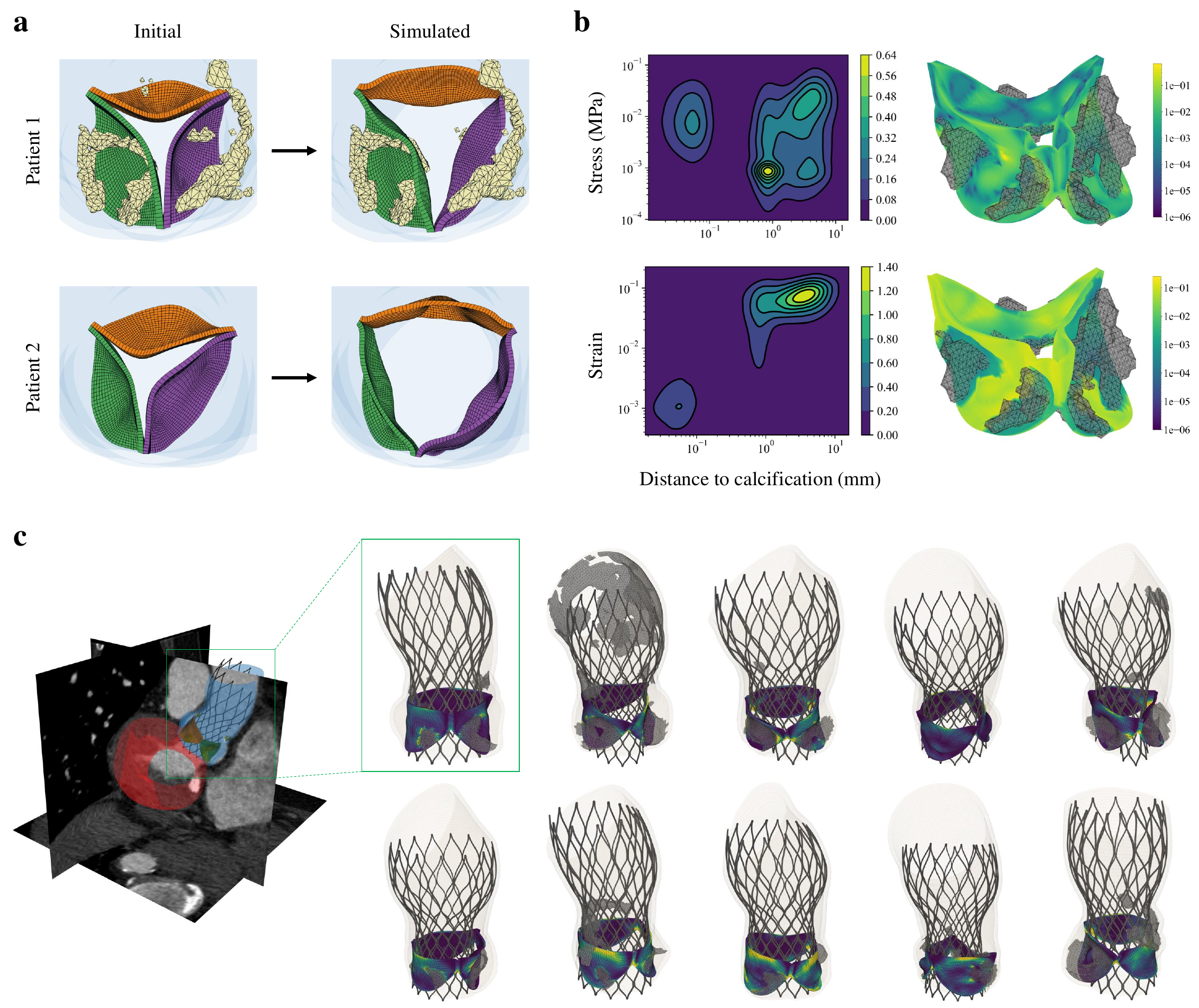}}
    \caption{(a) Valve opening simulations demonstrate the effects of calcification on the final leaflet positions. Yellow is the ground-truth calcification, left is the input valve geometry predicted by DeepCarve, and right is the deformed geometry after finite element analysis. Movement is clearly restricted near calcified regions. (b) Stress (top) and strain (bottom) analyses from valve opening simulations. Left is the Gaussian KDE plot of stress/strain vs. distance to calcification from the aggregate of 35 test-set patient simulations. Right is one test-set patient with stress/strain overlaid with the valve leaflets, plus the ground-truth calcification (gray) for reference. (c) TAVR stent deployment simulation results. Left: image and simulated geometry overlay. Right: maximum principal stress magnitudes plotted on the aortic valve leaflets for 10 different test-set patients.}
    \label{fig:combined_simulations}
\end{figure*}

We performed two sets of solid mechanics simulations to assess different aspects of our final C-MAC meshes. First is valve opening simulations, which demonstrate our method's unique ability to set up large-scale simulations. Second is TAVR stent deployment simulations, which demonstrate our meshes' robustness to complex simulation environments (Fig.~\ref{fig:combined_simulations}). Simulations were performed using Abaqus (3DS, Dassault Systéms, Paris, France).

For valve opening, a uniform pressure of 8 mmHg was applied to the ventricular surface of the leaflets. The pressure was increased compared to previous work \citep{martin2015comparison} to represent elevated aortic valve pressure gradient in mild aortic stenosis. Valve opening simulations were performed across all 35 patients with \textit{fully automated} simulation setups. This was possible because we maintained the mesh correspondence of heart meshes, and all boundary and loading conditions were defined with respect to the heart mesh elements. This allowed us to both qualitatively confirm the consistency of our results across a large number of samples, and also quantify our observations across the entire test-set population.

For TAVR stent deployment, we modeled a 26 mm, self-expandable, first-generation Medtronic CoreValve device (Medtronic, Minneapolis, MN, USA) following previous works \citep{mao2018numerical,caballero2020impact}. We selected 10 patient-specific anatomies that are suitable for this device size according to the manufacturer’s recommendation of CoreValve system \citep{medtronic2017corevalve}. Briefly, the crimped TAVR stent with an exterior diamater of 6mm \citep{martin2015comparison} was aligned coaxially within the aortic root and centered into the aortic annulus, following the manufacturer’s recommendations \citep{medtronic2017corevalve}. The stent was deployed inside the aortic root by axially moving the cylindrical sheath towards the ascending aorta.

From both simulations, we extracted two main pieces of information. First is the effect of the calcification on the leaflet movement. The leaflet movement is clearly restricted near the calcified regions (Fig.~\ref{fig:combined_simulations}a). This can be attributed to the high stiffness of the calcification and the aortic wall, which incur both intrinsic resistance from the leaflet elements and external push-back from the physical connection to the aortic wall. Second set of information was the stress/strain values. The stress/strain values are significantly influenced by the location of the calcification. The simulations were performed using completely auto-generated meshes from DeepCarve and C-MAC, and the stress/strain results correspond well to the location of the ground-truth calcification.

For valve opening, we furhter performed Gaussian kernel density estimation (KDE) \citep{scott2015multivariate,2020SciPy-NMeth} on the aggregated data points of all test-set simulations to estimate the joint distribution of the leaflet nodes' stress/strain values vs. their distances to the nearest calcification (Fig.~\ref{fig:combined_simulations}b). The KDE plots (1) confirm our qualitative observations that nodes closer to calcification have low strain, and (2) reveal new insights that the leaflet stress may exhibit a complex pattern at 1-10 millimeters away from the calcification. More detailed analyses of the simulation results may lead to important insights on the growth mechanism of calcification as well as the intricate leaflet behavior associated with the positioning of the calcification within the aortic root.

For stent deployment, the key finding was the unmatched robustness of C-MAC's outputs in the complex multi-body simulation. Using the original calcification meshing algorithm from DeepCarve, 4 out of the 10 simulations failed due to calcification element distortions. Using C-MAC, all 10 simulations successfully ran to completion. The simulation results in Fig.~\ref{fig:combined_simulations}c concur with previous observations that the calcified regions are associated with higher levels of stress. This further confirms the high robustness of our meshing algorithm and the benefit it provides for complex downstream tasks.

\section*{Discussion}
Our extensive evaluations suggest that C-MAC is a robust fully automated solution for incorporating aortic calcification mesh into an existing heart mesh. (1) Our deep learning segmentation model provides a good initial calcification segmentation. (2) Our segmentation post-processing algorithm enhances the calcification-to-heart anatomical consistency while slightly compromising the segmentation volume accuracy. (3) Our meshing algorithm accurately converts the input voxelgrid segmentation into surface meshes with great element quality along both contact and non-contact surfaces. The meshing algorithm also preserves the original heart mesh topology.

Unlike the original DMTet paper \citep{shen2021deep}, we did not train a separate neural network to expedite the DMTet optimization process. Instead, we opted for a per-image optimization approach. The main reason for this choice was the long processing time of TetGen for the background mesh generation. Unlike \cite{shen2021deep} where the background mesh was a fixed lattice of regularly spaced tetrahedra, our meshing workflow requires a background mesh that contains the patient-specific heart geometry in the inner volume. Since the element density and configuration can change drastically based on the leaflet conformation, we could not reliably construct the background mesh without repeating TetGen for each scan separately. This poses a significant restriction on our training scheme since repeating TetGen for every augmented training sample would significantly increase the training time (from $\sim$1 second to $\sim$15 seconds per sample) or storage ($\sim$20 MB per background mesh for hundreds of thousands of samples).

Furthermore, even if we were able to successfully train a deep learning model for the DMTet optimization process, it would only reduce the total run-time from $\sim$1 minute to $\sim$40 seconds because the other major time-consuming processes have nothing to do with DMTet optimization. Therefore, we decided to simply use the proposed per-image optimization approach. This approach has an additional minor benefit that we can slightly randomize the output with different initialization for DMTet optimization if we are unsatisfied with our output.


Our current approach for segmentation post-processing has a trade-off between calcification segmentation accuracy and anatomical consistency because we fix DeepCarve's heart structures while extending the calcification segments to remove the spatial gaps. We aimed to achieve the best possible solution for this approach using an adaptive closing kernel, but the inherent reduction in volume overlap is likely unavoidable. An interesting future direction would be to jointly optimize the heart structures and the calcification segments to deform the heart structures towards the predicted calcification, instead of simply extending the calcification. This approach is technically more challenging because it requires simultaneous optimization of multiple metrics, such as the calcification accuracy, heart surface accuracy, heart mesh element quality, and also requires a robust formulation for the selective attraction between only the contact surfaces of the heart and calcification meshes.

Another important consideration for C-MAC is the usability of the final meshes. Since we optimize for good element quality using DMTet optimization, the key remaining consideration is the manifoldness of the output. One of the main strengths of marching tetrahedra is its ability to generate topologically consistent outputs due to its unambiguous tessellation of the tetrahedral lattice. The same principle applies to the raw non-optimized DMTet output with our custom background mesh and calcification segmentation, which means the output should be manifold surfaces without any modifications. The main issue with the non-optimized DMTet output is that TetGen can occasionally fail due to the highly irregular triangular elements from the raw node-sampled SDF.

We proposed DMTet optimization to maintain the original manifoldness of DMTet while improving the element quality. The element quality improvements are straightforwardly obtained using $\mathcal{L}_{DMTet}$ (Eq.~\ref{eq:DMTet_loss}), which optimizes for various desired triangle properties. For the manifoldness, we first argue that \FuncCall{ModifyEdgeCases} (Eq.~\ref{eq:modified_DMTet_SDF}) does not alter the manifoldness of the original DMTet output. From Eq.~\ref{eq:modified_v_ab}: (1) For any tetrahedron with an SDF sign-change across the fake node, the output surface will always be a heart surface element. DeepCarve's heart meshes all have manifold surfaces by design, so the DMTet output at contact surfaces will also be manifold. (2) For any tetrahedron with an SDF sign-change across a boundary node, the output surface will always only \textit{expand} its borders along the boundary surface, and the amount of expansion will at worst cause coincident faces. This means no self-intersections will be introduced by our modification, so the DMTet output will be manifold.

For \FuncCall{DMTetOpt}'s $\delta$ optimization (Alg.~\ref{alg:dmtet_opt}), $\Delta SDF$ is constrained to only change each nodal SDF's magnitude while keeping its original sign, which prevents any topological changes. The only remaining potential concern is $\Delta V_{bg}$. Given that $\mathcal{L}_{DMTet}$ is only calculated for tetrahedra with nodal SDF sign-changes, the $\Delta V_{bg}$ optimization will also only be applied to those elements' nodes. In this case, a potential way for $\Delta V_{bg}$ to cause non-manifoldness is to cause self-intersections by heavily displacing the background mesh nodes near the segmentation isosurface. This is strongly discouraged by many of our design choices, such as (1) $\mathcal{L}_{DMTet}$'s surface smoothness constraints and $\Delta V_{bg}$ regularization and (2) our strategy of updating the nodal SDF at the new node coordinate for each \FuncCall{DMTetOpt} iteration. Empirically, $\FuncCall{DMTetOpt}$ consistently generates high quality manifold meshes, leadig to a TetGen success rate of 100\%.


Clinical validation of solid mechanics simulations is extremely challenging because post-TAVR CT's are not routinely collected and the ground-truth stress/strain values for \textit{in vivo} anatomical structures are nearly impossible to obtain. The solid mechanics simulations in this work instead illustrate two other important points: (1) our generated meshes can handle complex solid interactions such as TAVR stent deployments and (2) the high-quality mesh output with mesh correspondence allows for large-batch simulations, even with the addition of calcification. Further clinical validation of our simulations will be performed with flow-based simulations in future studies using the more routinely collected Doppler echo data.

In this work, we propose an automated end-to-end image-to-mesh solution for incorporating calcification into an existing heart geometry. Our solution includes (1) a segmentation algorithm that encourages anatomical consistency between the predicted calcification and heart mesh and (2) a meshing algorithm that establishes complete contact surface matching between the two geometries while maintaining the original heart mesh topology. Our technique allows for large-batch simulations due to the preservation of inter-patient mesh correspondence of the heart geometry. We demonstrated the viability of such downstream tasks using solid mechanics simulations. Our method may help accelerate the development and usage of complex physics-driven simulations for cardiovascular applications.





 \label{sec:discussion}

\section*{Methods}
\subsection*{Data acquisition and preprocessing} \label{sec:data}

We used the same dataset as DeepCarve with some minor modifications. The original dataset consisted of 80 CTA scans and the corresponding labels for the heart structures and calcification. 14 of the 80 CT scans were from 14 different patients in the training set of the MM-WHS public dataset \cite{zhuang2016multi}. The remaining 66 scans were from 55 IRB-approved TAVR patients at Hartford hospital. We used more than one time point for some of the Hartford patients. Of the 80 total scans, we used 35/10/35 scans for training/validation/testing.  We ensured that the testing set had no overlapping patients with the training/validation sets. All evaluations were performed with the 35 test-set cases. All patients had tricuspid aortic valves and varying levels of calcification. Some scans had no aortic calcification.

The calcification segmentation was obtained using adaptive thresholding and manual post-processing. The original dataset included calcification that is within some distance away from any of the pointclouds. For this work, we further processed it to only include segments around the aorta. This was mainly to omit calcification along the mitral valve, which would require accurate mitral valve leaflet geometry for us to properly apply our meshing techniques.

Both Gbd and Gch require extra training labels extracted from the original calcification segmentation. Gbd requires voxelgrid SDF, which we obtained using isosurface extraction and point-to-mesh distance from each voxel position. Gch requires calcification pointclouds, which we obtained using isosurface extraction and surface point sampling.

For preprocessing, we rescaled all images and labels to an isotropic spatial resolution of 1.25mm$^3$ and cropped them to 128$^3$ voxels. The default crop center was the center of the ground-truth heart pointclouds' bounding box. For translation augmentation, the crop center was offset by a random sample from a 3D Gaussian (stdev = crop\_width/3), and the offset amount was capped to prevent any labeled structures being outside the cropped region. The images were not pre-aligned with any additional transforms.

Image intensities were first clipped to [lower bound, upper bound], and further min-max normalized to [0,1]. The lower and upper bounds were fixed for DeepCarve at [-158,864] Hounsefield Units (HU) and chosen between three different ranges for calcification segmentation, as specified in the results tables (Table.~\ref{table:results_dl_seg}).




\subsection*{Deep learning model training details} \label{sec:method_dl_seg}





Dice (a.k.a. DSC) \citep{drozdzal2016importance,sudre2017generalised} and cross entropy (CE) \citep{ronneberger2015u} are commonly used for training segmentation models. Dice has several advantages over CE, such as being more robust to unbalanced targets and being able to directly optimize a common evaluation metric, but it is ineffective at handling empty targets. A common workaround is to instead optimize the combined DiceCE loss \citep{ma2021loss}, which can be defined for binary segmentation tasks as



\begin{align}
\begin{split}
    \mathcal{L}_{DiceCE}(y, \hat{y}) \;=\; & 1 - 2 \frac{\sum_{i=1}^N y_i \hat{y}_i}{\sum_{i=1}^N y_i + \sum_{i=1}^N \hat{y}_i + \epsilon} \\
    + \; &\lambda \left[ \frac{1}{N} \left( \sum_{i=1}^N y_i \log(\hat{y}_i) + (1-y_i) \log(1-\hat{y}_i)\right) \right]
\end{split}
\end{align}

\noindent where $y_i$ is the target and $\hat{y}_i$ is the predicted value at each $i^{\textrm{th}}$ voxel, $\lambda$ is a weighting hyperparameter, and $\epsilon$ is an error term that prevents division by 0. To prevent numerical issues with $\log(0)$, the $\log$ output is set to have a minimum bound at -100 \citep{paszke2017automatic}.

Generalized Dice Loss (GDL) is a variation of the Dice loss that handles multi-class segmentation with a per-class weighting term that mitigates biases towards larger segments \citep{sudre2017generalised}. For a binary segmentation task, we can treat the background as its own class, in which case

\begin{equation}
    \mathcal{L}_{GDL}(y, \hat{y}) = 1 - 2 \frac{\sum_{c=1}^2 w_c \sum_i y_{ci} \hat{y}_{ci}}{\sum_{c=1}^2 w_c \sum_i (y_{ci} + \hat{y}_{ci})}
\end{equation}

\noindent where $w_c = \dfrac{1}{(\sum_i y_{ci} + \epsilon_c)^2}$. We set $\epsilon_0 = -100$ and $\epsilon_1 = 100$ via hyperparameter tuning, where $c=0$ is background and $c=1$ is calcification. The $\epsilon_c$ helps prevent division by 0 and helps maintain a smooth Dice-like behavior for both empty and sparse targets. Empirically, GDL performed better than DiceCE for our task, so it was chosen as the main region-based loss for the rest of our experiments.


All variants of distribution-based losses (i.e. CE) and region-based losses (i.e. Dice) penalize each voxel prediction independently, and fail to capture the magnitude of error from the original segmentation boundary. The boundary loss was proposed to combat this phenomenon \citep{kervadec2019boundary}:

\begin{equation}
    \mathcal{L}_{Gbd}(y, \hat{y}) = \mathcal{L}_{GDL}(y, \hat{y}) + \lambda \sum_i -SDF(y)_i \hat{y}_i
\end{equation}

\noindent where $SDF(y)$ is the signed distance function (SDF) calculated using the ground-truth segmentation $y$. By this definition, SDF is assumed to be positive on the inside and negative on the outside of the $y=0.5$ isosurface. The weighting hyperparameter $\lambda$ is adjusted based on training epochs, similar to \cite{kervadec2019boundary}. In our case, $\lambda=0$ at epoch 0 and was linearly increased to $\lambda=1000$ from epochs 1000-2000. We clipped the SDF from -3 to 3 for numerical stability.

Finally, instead of relying on the integral approximation approach as did \cite{kervadec2019boundary}, we evaluated the effectiveness of directly optimizing for the symmetric chamfer distance \citep{wang2018pixel2mesh} between the predicted and target segmentations:

\begin{align} \label{eq:chamfer}
\begin{split}
    \mathcal{L}_{Gch}&(y, \hat{y}) \;=\; \mathcal{L}_{GDL}(y, \hat{y}) \\
    &+ \lambda \left( \frac{1}{\mid A \mid}  \sum_{a \in A} \min_{b \in B} \norm{a-b}_2^2 
    + \frac{1}{\mid B \mid} \sum_{b \in B} \min_{a \in A} \norm{b-a}_2^2 \right)
\end{split}
\end{align}

\noindent where $A$ and $B$ are pointclouds sampled on the segmentations' extracted surfaces. We used DMTet to extract the surface from the predicted segmentation, which makes the operation differentiable with respect to the segmentation model parameters \citep{shen2021deep}. $\lambda = 0$ at epoch 0 and linearly increased to $\lambda=0.1$ from epochs 1000-2000.

All models were trained for 4000 epochs with the Adam optimizer, at a learning rate of 1e-4. For data augmentation, we used random translation and b-spline deformation.

We also compared our models' performance against nnU-net, a popular self-configuring DL segmentation method. For fair comparisons, we trained the nnU-net with the same training/validation/test splits as our own models, which effectively discarded nnU-net's default ensembling operations.


\subsection*{Segmentation post-processing details} \label{sec:method_post_process}

\begin{algorithm}
    \caption{Post-processing of ca2 segmentation} \label{alg:seg_post_process}
    \begin{algorithmic}[1]
    \Function{PostProcessCa2Seg}{$y_{0}, M_{heart}$}
    \State $\{y_{i}\} \gets \FuncCall{GroupSegByLeaflets}(y_{0})$ \Comment{$i\in\{1,2,3\}$}
    \While {not satisfied}
        \For {$y_i \textrm{ in } \{y_{i}\}$}
            \State $\hat{y}_{heart} \gets \FuncCall{FilterHeartSeg}(y_i, M_{heart})$
            \State $y_i \gets \FuncCall{AdaptiveClose}(y_i, \hat{S}_{heart})$
        \EndFor
    \EndWhile
    \State $y_{ca2} \gets \FuncCall{SubtractAndFilterSeg}(\{y_i\}, M_{heart})$
    \State \Return{$y_{ca2}$}
    \EndFunction
    \end{algorithmic}
    \hrulealg
    $S_i$ is assigned in-place, so ${S_i}$ is updated with each iteration.
\end{algorithm}

(\FuncCall{GroupSegByLeaflets}) First, we split the calcification segmentation into three groups - one group for each aortic valve leaflet. Each island of the segmentation was assigned to the closest aortic valve leaflet, measured by the mean symmetric chamfer distance.


(\FuncCall{FilterHeartSeg}) Using the grouped calcification segments and the predicted heart mesh, we extracted heart segments that are within some distance away from the calcification. We first converted the heart mesh to heart segments using the image stencil operation \citep{schroeder1998visualization}, and only kept the segments intersecting with dilated calcification segments. The dilation filter was 7x7x7 voxels and adaptive, meaning the filter values were ellipsoidal with the major principal axis pointing in the same direction as the closest heart mesh node's surface normal.


(\FuncCall{AdaptiveClose}) We used a two-step morphological close operations on the grouped calcification segments and the corresponding filtered heart segments. The first closing operation was performed with the same adaptive kernel described in \FuncCall{FilterHeartSeg}, and the second closing operation was performed with an isotropic spherical 3x3x3 filter.

(\FuncCall{SubtractAndFilterSeg}) Due to the limited resolution of the voxelgrid representation, directly meshing from the combined segments results in many undesired protrusions in the final mesh. To avoid this, we filtered the combined segments by subtracting away the heart segments and performing volume-based island removals. For the subtraction, we tripled the segmentation resolution to make sure that we can capture the thin leaflet geometries in the voxelgrid representation. The final segmentation for downstream meshing ($y_{ca2}$) therefore has a 3x voxelgrid resolution compared to the original image.







\subsection*{Background mesh generation details} \label{sec:mesh_bg}


\begin{algorithm}
    \caption{Background mesh generation} \label{alg:mesh_bg}
    \begin{algorithmic}
    \Function{GenerateBackgroundMesh}{$M_{heart}$}
    \State $S_{aorta} \gets \FuncCall{ExtractAortaSurf}(M_{heart})$
    \State $S_{offset} \gets \FuncCall{OffsetSurf}(M_{heart})$
    \State $M_{prelim} \gets \FuncCall{TetGenAndHollow}(S_{aorta}, S_{offset})$
    \State $M_{bg} \gets \FuncCall{CreateFakeElems}(M_{prelim})$
    \State \Return{$M_{bg}$}
    \EndFunction
    \end{algorithmic}
\end{algorithm}

(\FuncCall{ExtractAortaSurf}) Since we were focusing on aortic calcification, the original heart surface consisted of the surface elements of the aorta and the aortic valve leaflets, which we can extract using standard meshing libraries \citep{schroeder1998visualization}.

(\FuncCall{OffsetSurf}) For the outer bound of the background mesh, we generated an offset surface that includes all areas within 10 voxel spacings away from the aortic surface. The extracted aortic surface and the offset surface were combined by addition, which was easily doable because the offset surface and the extracted heart surface are clean manifold surfaces with no intersections by construction.

(\FuncCall{TetGenAndHollow}) The merged surfaces were processed by TetGen for constrained tetrahedral meshing, and then the elements inside the aortic surface were removed.

(\FuncCall{CreateFakeElems}) Lastly, we created ``fake" tetrahedral elements by adding a ``fake" node to all background boundary surfaces, i.e. the aortic surface and the offset surface elements.

\FuncCall{TetGenAndHollow} and \FuncCall{CreateFakeElems} are crucial for accurate meshing. The rationale will be explained further in the following subsection*.


\subsection*{DMTet optimization details}


\begin{algorithm}
    \caption{DMTet optimization} \label{alg:dmtet_opt}
    \begin{algorithmic}
    \Function{DMTetOpt}{$y_{ca2}, M_{bg}$}
    \State $(V_{bg}, E_{bg}) \gets M_{bg}$
    \State $\delta \sim \mathcal{U}_n(-1e-3, 1e-3)$
    \For{$n_{opt}$}
        \State $(\Delta V_{bg}, \Delta SDF) \gets \delta$
        \State $\widetilde{V}_{bg} \gets V_{bg} + \Delta V_{bg}$
        \State $\widetilde{M}_{bg} \gets (\widetilde{V}_{bg}, E_{bg})$
        \State $SDF \gets \FuncCall{LinearMapAndInterp}(y_{ca2}, \widetilde{V}_{bg})$
        \State $SDF \gets \FuncCall{ModifyEdgeCases}(SDF)$ \Comment{Eq.~\ref{eq:modified_DMTet_SDF}}
        \State $\widetilde{SDF} \gets SDF + \Delta SDF$
        \State $S_{DMTet} \gets \FuncCall{DMTet}(\widetilde{SDF}, \widetilde{M}_{bg})$ \Comment{Eq.~\ref{eq:dmtet_v_ab}}
        \State $(V, E) \gets S_{DMTet}$
        \State $\delta \gets \FuncCall{Adam}(\delta, \mathcal{L}_{DMTet}(V, E, \Delta V_{bg}))$ \Comment{Eq.~\ref{eq:DMTet_loss}}
    \EndFor
    \State \Return{$S_{DMTet}, \mathcal{L}_{DMTet}, \delta$}
    \EndFunction
    \end{algorithmic}
\end{algorithm}

(\FuncCall{LinearMapAndInterp}) We first converted the predicted segmentation to a simplified voxelgrid SDF via linear mapping $SDF(\hat{y}) = 2\hat{y}-1$, which means $\hat{y}_i \in \{0,1\} \rightarrow SDF(\hat{y})_i \in \{-1,1\}$. Then, we trilinearly interpolated the SDF at each background mesh node.

To demonstrate the need for further processing, let us consider applying marching tetrahedra with the default $SDF_i \in [-1,1]$ and the preliminary background mesh. Due to the smooth transition of the nodal SDF magnitudes, \textit{none} of the resulting surface mesh would match exactly with the original background mesh nodes. This is counterproductive, as our desired output is a surface mesh with coincident nodes and edges along the contact surfaces.

(\FuncCall{ModifyEdgeCases}) To enforce complete contact surface matching, we make two modifications to the nodal SDF after the initial interpolation:

\begin{equation} \label{eq:modified_DMTet_SDF}
    SDF_i = 
    \begin{dcases*}
    0 & if background boundary node \\
    -1e12 & if fake node \\
    SDF_i & else
    \end{dcases*}
\end{equation}

\noindent Then, from Eq.~\ref{eq:dmtet_v_ab}, this results in the following node assignments:

\begin{equation} \label{eq:modified_v_ab}
    v_{ab}^{\prime} \approx 
    \begin{dcases*}
    v_a & if background boundary node \\
    v_b & if fake node \\
    v_{ab}^{\prime} & else
    \end{dcases*}
\end{equation}

\noindent Recall that (1) all elements inside the heart volume were removed via $\FuncCall{TetGenAndHollow}$ and (2) the fake node was added to the background boundary surface to generate fake tetrahedral elements via $\FuncCall{CreateFakeElems}$. Together with Eq.~\ref{eq:modified_v_ab}, this means that any SDF sign change involving a boundary node will \textit{always} result in that node being included as a part of the DMTet output. This is exactly our intended output with complete contact surface matching with the original boundary elements.


Applying Eq.~\ref{eq:modified_DMTet_SDF} is enough for establishing complete contact surface matching. However, another known problem of marching tetrahedra is the irregular elements of the output surface \citep{treece1999regularised}. Although some related solutions exist, our task also requires that the contact surfaces be preserved during mesh processing. DMTet optimization is the first step of our two-part solution. Constrained surface remeshing is the optional second step (Section~\ref{sec:method_remesh}).

Similar to the original DMTet \citep{shen2021deep}, we optimized (1) the deformation of the background mesh and (2) the offset from the initial interpolated nodal SDF. Unlike the original paper, we do not train a deep learning model for the prediction. Instead, we perform the optimization for each inference target independently. The rationale for this choice is described in Section~\ref{sec:discussion}. During the optimization process, we minimize the output's deviation from the original segmentation by penalizing the background mesh deformation and prohibiting sign changes in the nodal SDF.

The overall idea is to improve the mesh quality while minimizing its deviation from the original segmentation. The loss is a combination of Laplacian smoothing, edge length, edge angle, and deformation penalty (Eq.~\ref{eq:DMTet_loss}). The first three losses help improve the overall mesh and individual element qualities, while the deformation penalty helps minimize surface shrinking. Eq.~\ref{eq:modified_DMTet_SDF} is applied after each optimization step, so only the non-contact nodes' SDF values are optimized during this process.

\begin{align} \label{eq:DMTet_loss}
\begin{split}
    \mathcal{L}_{DMTet}(V, E, \Delta V_{bg}) =\; & \lambda_0 \sum_{v_i \in V}  \norm{\frac{1}{\mid \mathcal{N}(v_i) \mid} \sum_{v_j \in \mathcal{N}(v_i)} v_i - v_j}_2 \\
    +\; & \lambda_1 \left[ \sum_{v_i \in V} \sum_{v_j \in \mathcal{N}(v_i)} \left(\norm{v_i - v_j}_2 -\epsilon \right)^2 \right]^{\frac{1}{2}} \\
    +\; & \lambda_2 \left[ \sum_{a \in A(V,E)} (a-\alpha)^2 * \sigma(a) \right]^{\frac{1}{2}} \\
    +\; & \lambda_3 \sum_{v_i \in V_{bg}} \norm{\Delta v_i}_2  \\
\end{split}
\end{align}

\noindent (V,E): vertices and edges of the DMTet output. $\mathcal{N}$: neighboring nodes. $A$: angles between all edges. $\epsilon$: desired edge length, set to 0.1 for our task. $\alpha$: desired edge angle, set to 60 degrees. $\lambda_i$: hyperparameters, each set to [10, 1, 1, 0.5]. $\sigma(a) = (1-sigmoid(a-10)) + sigmoid(150)$: filter to minimize the penalty for ``good enough" angles, i.e. between 10 and 150 degrees.

We used the Adam optimizer with a learning rate of 1e-2 and achieved convergence at around $n_{opt} = 100$. For the final DMTet output, we cleaned the mesh by removing overlapping nodes (distance $<$ 1e-2) and elements with repeated nodes (i.e. points and edges), which are caused by Eq.~\ref{eq:modified_DMTet_SDF}.

\subsection*{Constrained surface remeshing details} \label{sec:method_remesh}

Although the DMTet output is already a suitable manifold surface for the final tetrahedralization, it can be further refined via surface remeshing (Alg.~\ref{alg:remesh}). The main remeshing algorithm we used is ACVD, a Voronoi diagram-based vertex clustering method \citep{valette2008generic}.

\begin{algorithm}
    \caption{Constrained remeshing by vertex-clustering} \label{alg:remesh}
    \begin{algorithmic}
    \Function{ConstrainedRemesh}{$S_{ca2}$}
    \For{$n_{remesh}$}
        \State ${S_{contact}, S_{free}} \gets \FuncCall{SeparateContactSurf}(S_{ca2})$
        \State $S_{free2} \gets \FuncCall{ConstrainedClustering}(S_{free})$
        \State $S_{ca2} \gets \FuncCall{Merge}(S_{contact}, S_{free2})$
    \EndFor
    \State \Return{$S_{ca2}$}
    \EndFunction
    \end{algorithmic}
\end{algorithm}

To preserve the contact surface elements, we made two modifications to the remeshing steps.

(\FuncCall{SeparateContactSurf}) First, we split the initial surface into contact and non-contact elements by checking the number of nodes merged with the original heart mesh's nodes. If all three nodes of a triangular element are coincident with heart mesh nodes, then that element is a contact element.

(\FuncCall{ConstrainedClustering}) We performed the initial ACVD clustering step on the non-contact surface, which assigns a cluster index to each node. The number of initial clusters was set to 80\% of the number of nodes. The nodes belonging to both the contact and non-contact elements were preserved by assigning new unique cluster indices. Note that the percentage of vertices actually being clustered progressively decreases with each remeshing step because of our cluster re-assigning scheme. Finally, we performed the standard ACVD triangulation to obtain the remeshed non-contact surface.

(\FuncCall{Merge}) We merged the contact surfaces and the remeshed non-contact surfaces to obtain the final output surface mesh. This step was easy to perform because we preserved the coincident nodes between the contact and non-contact surfaces using constrained clustering.

The constrained remeshing was repeated as many times as necessary to achieve the desired element size and density. We often stopped at $n_{remesh} = 15$ to prevent oversimplification. The remeshing step is optional, as we can obtain good quality tetrahedral calcification meshes directly from the output of \FuncCall{DMTetOpt}. However, we found that the remeshing step generally further improves the element quality for downstream applications.

\subsection*{Final tetrahedralization} \label{sec:final_tet}

We obtained the final tetrahedral calcification mesh by applying TetGen on the surface mesh from the previous subsections. This step was straightforward because we designed the surface meshes to be clean and manifold. For node-stitching, we first assigned each connected component of calcification to a leaflet based on the lowest mean symmetric chamfer distance, and nodes from the assigned leaflet or the aortic wall were merged with calcification nodes if the distances were less than 1e-3. Connected components of calcification that had no merged nodes with heart surfaces were removed.

\subsection*{Statistical analyses}

The scipy package \citep{2020SciPy-NMeth} was used to perform related-sample t-tests with two-sided alternative hypothesis. Metrics from repeated meshing trials were first averaged across each patient.

\section*{Acknowledgments}
This work was supported by the National Heart, Lung, and Blood Institute (NHLBI) of the National Institute of Health (NIH), grant F31HL162505.

\section*{Author contributions}

DHP designed the algorithm and performed all experiments other than the finite element simulations. ML performed the finite element simulations. DHP and ML wrote the manuscript. TK provided guidance for algorithm assessment. DHP, ML, RG, and JSD reviewed the results. CO and ER evaluated the outputs. DHP and RM curated the dataset.

\section*{Competing interests}
A provisional patent application No. 63/611,903 has been filed with the contents of this manuscript. Inventor: DHP.


\bibliographystyle{unsrtnat}
\bibliography{refs}

\begin{thebibliography}{57}
\providecommand{\natexlab}[1]{#1}
\providecommand{\url}[1]{\texttt{#1}}
\expandafter\ifx\csname urlstyle\endcsname\relax
  \providecommand{\doi}[1]{doi: #1}\else
  \providecommand{\doi}{doi: \begingroup \urlstyle{rm}\Url}\fi

\bibitem[Greenland et~al.(2004)Greenland, LaBree, Azen, Doherty, and Detrano]{greenland2004coronary}
Philip Greenland, Laurie LaBree, Stanley~P Azen, Terence~M Doherty, and Robert~C Detrano.
\newblock Coronary artery calcium score combined with framingham score for risk prediction in asymptomatic individuals.
\newblock \emph{Jama}, 291\penalty0 (2):\penalty0 210--215, 2004.

\bibitem[Chen et~al.(2017)Chen, Budoff, Reilly, Yang, Rosas, Rahman, Zhang, Roy, Lustigova, Nessel, et~al.]{chen2017coronary}
Jing Chen, Matthew~J Budoff, Muredach~P Reilly, Wei Yang, Sylvia~E Rosas, Mahboob Rahman, Xiaoming Zhang, Jason~A Roy, Eva Lustigova, Lisa Nessel, et~al.
\newblock Coronary artery calcification and risk of cardiovascular disease and death among patients with chronic kidney disease.
\newblock \emph{JAMA cardiology}, 2\penalty0 (6):\penalty0 635--643, 2017.

\bibitem[Witteman et~al.(1986)Witteman, Kok, Van~Saase, and Valkenburg]{witteman1986aortic}
JacquelineC~M Witteman, FransJ Kok, JanL~CM Van~Saase, and HansA Valkenburg.
\newblock Aortic calcification as a predictor of cardiovascular mortality.
\newblock \emph{The Lancet}, 328\penalty0 (8516):\penalty0 1120--1122, 1986.

\bibitem[Nicoll and Henein(2014)]{nicoll2014predictive}
Rachel Nicoll and Michael~Y Henein.
\newblock The predictive value of arterial and valvular calcification for mortality and cardiovascular events.
\newblock \emph{IJC Heart \& Vessels}, 3:\penalty0 1--5, 2014.

\bibitem[Sangiorgi et~al.(1998)Sangiorgi, Rumberger, Severson, Edwards, Gregoire, Fitzpatrick, and Schwartz]{sangiorgi1998arterial}
Giuseppe Sangiorgi, John~A Rumberger, Arlen Severson, William~D Edwards, Jean Gregoire, Lorraine~A Fitzpatrick, and Robert~S Schwartz.
\newblock Arterial calcification and not lumen stenosis is highly correlated with atherosclerotic plaque burden in humans: a histologic study of 723 coronary artery segments using nondecalcifying methodology.
\newblock \emph{Journal of the American College of Cardiology}, 31\penalty0 (1):\penalty0 126--133, 1998.

\bibitem[Durham et~al.(2018)Durham, Speer, Scatena, Giachelli, and Shanahan]{durham2018role}
Andrew~L Durham, Mei~Y Speer, Marta Scatena, Cecilia~M Giachelli, and Catherine~M Shanahan.
\newblock Role of smooth muscle cells in vascular calcification: implications in atherosclerosis and arterial stiffness.
\newblock \emph{Cardiovascular research}, 114\penalty0 (4):\penalty0 590--600, 2018.

\bibitem[Mohler(2004)]{mohler2004mechanisms}
Emile~R Mohler.
\newblock Mechanisms of aortic valve calcification.
\newblock \emph{American Journal of Cardiology}, 94\penalty0 (11):\penalty0 1396--1402, 2004.

\bibitem[Pawade et~al.(2019)Pawade, Sheth, Guzzetti, Dweck, and Clavel]{pawade2019and}
Tania Pawade, Tej Sheth, Ezequiel Guzzetti, Marc~R Dweck, and Marie-Annick Clavel.
\newblock Why and how to measure aortic valve calcification in patients with aortic stenosis.
\newblock \emph{JACC: Cardiovascular Imaging}, 12\penalty0 (9):\penalty0 1835--1848, 2019.

\bibitem[Ge and Sotiropoulos(2010)]{ge2010direction}
Liang Ge and Fotis Sotiropoulos.
\newblock Direction and magnitude of blood flow shear stresses on the leaflets of aortic valves: is there a link with valve calcification?
\newblock 2010.

\bibitem[Halevi et~al.(2016)Halevi, Hamdan, Marom, Lavon, Ben-Zekry, Raanani, Bluestein, and Haj-Ali]{halevi2016fluid}
Rotem Halevi, Ashraf Hamdan, Gil Marom, Karin Lavon, Sagit Ben-Zekry, Ehud Raanani, Danny Bluestein, and Rami Haj-Ali.
\newblock Fluid--structure interaction modeling of calcific aortic valve disease using patient-specific three-dimensional calcification scans.
\newblock \emph{Medical \& biological engineering \& computing}, 54:\penalty0 1683--1694, 2016.

\bibitem[Weinberg et~al.(2010)Weinberg, Mack, Schoen, Garc{\'\i}a-Carde{\~n}a, and Kaazempur~Mofrad]{weinberg2010hemodynamic}
Eli~J Weinberg, Peter~J Mack, Frederick~J Schoen, Guillermo Garc{\'\i}a-Carde{\~n}a, and Mohammad~R Kaazempur~Mofrad.
\newblock Hemodynamic environments from opposing sides of human aortic valve leaflets evoke distinct endothelial phenotypes in vitro.
\newblock \emph{Cardiovascular engineering}, 10:\penalty0 5--11, 2010.

\bibitem[Arzani and Mofrad(2017)]{arzani2017strain}
Amirhossein Arzani and Mohammad~RK Mofrad.
\newblock A strain-based finite element model for calcification progression in aortic valves.
\newblock \emph{Journal of biomechanics}, 65:\penalty0 216--220, 2017.

\bibitem[Qin et~al.(2020)Qin, Caballero, Mao, Barrett, Kamioka, Lerakis, and Sun]{qin2020role}
Tongran Qin, Andr{\'e}s Caballero, Wenbin Mao, Brian Barrett, Norihiko Kamioka, Stamatios Lerakis, and Wei Sun.
\newblock The role of stress concentration in calcified bicuspid aortic valve.
\newblock \emph{Journal of the Royal Society Interface}, 17\penalty0 (167):\penalty0 20190893, 2020.

\bibitem[Milhorini~Pio et~al.(2020)Milhorini~Pio, Bax, and Delgado]{milhorini2020valvular}
Stephan Milhorini~Pio, Jeroen Bax, and Victoria Delgado.
\newblock How valvular calcification can affect the outcomes of transcatheter aortic valve implantation.
\newblock \emph{Expert review of medical devices}, 17\penalty0 (8):\penalty0 773--784, 2020.

\bibitem[Pollari et~al.(2020)Pollari, Hitzl, Vogt, Cuomo, Schwab, S{\"o}hn, Kalisnik, Langhammer, Bertsch, Fischlein, et~al.]{pollari2020aortic}
Francesco Pollari, Wolfgang Hitzl, Ferdinand Vogt, Michela Cuomo, Johannes Schwab, Claudius S{\"o}hn, Jurij~M Kalisnik, Christian Langhammer, Thomas Bertsch, Theodor Fischlein, et~al.
\newblock Aortic valve calcification as a risk factor for major complications and reduced survival after transcatheter replacement.
\newblock \emph{Journal of cardiovascular computed tomography}, 14\penalty0 (4):\penalty0 307--313, 2020.

\bibitem[Wang et~al.(2015)Wang, Kodali, Primiano, and Sun]{wang2015simulations}
Qian Wang, Susheel Kodali, Charles Primiano, and Wei Sun.
\newblock Simulations of transcatheter aortic valve implantation: implications for aortic root rupture.
\newblock \emph{Biomechanics and modeling in mechanobiology}, 14:\penalty0 29--38, 2015.

\bibitem[Sturla et~al.(2016)Sturla, Ronzoni, Vitali, Dimasi, Vismara, Preston-Maher, Burriesci, Votta, and Redaelli]{sturla2016impact}
Francesco Sturla, Mattia Ronzoni, Mattia Vitali, Annalisa Dimasi, Riccardo Vismara, Georgia Preston-Maher, Gaetano Burriesci, Emiliano Votta, and Alberto Redaelli.
\newblock Impact of different aortic valve calcification patterns on the outcome of transcatheter aortic valve implantation: a finite element study.
\newblock \emph{Journal of biomechanics}, 49\penalty0 (12):\penalty0 2520--2530, 2016.

\bibitem[Kurugol et~al.(2015)Kurugol, Come, Diaz, Ross, Kinney, Black-Shinn, Hokanson, Budoff, Washko, and San Jose~Estepar]{kurugol2015automated}
Sila Kurugol, Carolyn~E Come, Alejandro~A Diaz, James~C Ross, Greg~L Kinney, Jennifer~L Black-Shinn, John~E Hokanson, Matthew~J Budoff, George~R Washko, and Raul San Jose~Estepar.
\newblock Automated quantitative 3d analysis of aorta size, morphology, and mural calcification distributions.
\newblock \emph{Medical physics}, 42\penalty0 (9):\penalty0 5467--5478, 2015.

\bibitem[Mahabadi et~al.(2009)Mahabadi, Bamberg, Toepker, Schlett, Rogers, Nagurney, Brady, Hoffmann, and Truong]{mahabadi2009association}
Amir~A Mahabadi, Fabian Bamberg, Michael Toepker, Christopher~L Schlett, Ian~S Rogers, John~T Nagurney, Thomas~J Brady, Udo Hoffmann, and Quynh~A Truong.
\newblock Association of aortic valve calcification to the presence, extent, and composition of coronary artery plaque burden: from the rule out myocardial infarction using computer assisted tomography (romicat) trial.
\newblock \emph{American heart journal}, 158\penalty0 (4):\penalty0 562--568, 2009.

\bibitem[Alqahtani et~al.(2017)Alqahtani, Boczar, Kansal, Chan, Dwivedi, and Chow]{alqahtani2017quantifying}
Abdulrahman~M Alqahtani, Kevin~E Boczar, Vinay Kansal, Kwan Chan, Girish Dwivedi, and Benjamin~JW Chow.
\newblock Quantifying aortic valve calcification using coronary computed tomography angiography.
\newblock \emph{Journal of cardiovascular computed tomography}, 11\penalty0 (2):\penalty0 99--104, 2017.

\bibitem[Bettinger et~al.(2017)Bettinger, Khalique, Krepp, Hamid, Bae, Pulerwitz, Liao, Hahn, Vahl, Nazif, et~al.]{bettinger2017practical}
Nicolas Bettinger, Omar~K Khalique, Joseph~M Krepp, Nadira~B Hamid, David~J Bae, Todd~C Pulerwitz, Ming Liao, Rebecca~T Hahn, Torsten~P Vahl, Tamim~M Nazif, et~al.
\newblock Practical determination of aortic valve calcium volume score on contrast-enhanced computed tomography prior to transcatheter aortic valve replacement and impact on paravalvular regurgitation: elucidating optimal threshold cutoffs.
\newblock \emph{Journal of Cardiovascular Computed Tomography}, 11\penalty0 (4):\penalty0 302--308, 2017.

\bibitem[Vlastra et~al.(2019)Vlastra, van~den Boogert, Krommenhoek, Bronzwaer, Mutsaerts, Achterberg, Bron, Niessen, Majoie, Nederveen, et~al.]{vlastra2019aortic}
Wieneke Vlastra, Thomas~PW van~den Boogert, Thomas Krommenhoek, Anne-Sophie~GT Bronzwaer, Henk~JMM Mutsaerts, Hakim~C Achterberg, Esther~E Bron, Wiro~J Niessen, Charles~BLM Majoie, Aart~J Nederveen, et~al.
\newblock Aortic valve calcification volumes and chronic brain infarctions in patients undergoing transcatheter aortic valve implantation.
\newblock \emph{The International Journal of Cardiovascular Imaging}, 35:\penalty0 2123--2133, 2019.

\bibitem[Grbic et~al.(2013)Grbic, Mansi, Ionasec, Voigt, Houle, John, Schoebinger, Navab, and Comaniciu]{grbic2013image}
Sasa Grbic, Tommaso Mansi, Razvan Ionasec, Ingmar Voigt, Helene Houle, Matthias John, Max Schoebinger, Nassir Navab, and Dorin Comaniciu.
\newblock Image-based computational models for tavi planning: from ct images to implant deployment.
\newblock In \emph{Medical Image Computing and Computer-Assisted Intervention--MICCAI 2013: 16th International Conference, Nagoya, Japan, September 22-26, 2013, Proceedings, Part II 16}, pages 395--402. Springer, 2013.

\bibitem[Harbaoui et~al.(2016)Harbaoui, Montoy, Charles, Boussel, Liebgott, Girerd, Courand, and Lantelme]{harbaoui2016aorta}
Brahim Harbaoui, Mathieu Montoy, Paul Charles, Loic Boussel, Herv{\'e} Liebgott, Nicolas Girerd, Pierre-Yves Courand, and Pierre Lantelme.
\newblock Aorta calcification burden: towards an integrative predictor of cardiac outcome after transcatheter aortic valve implantation.
\newblock \emph{Atherosclerosis}, 246:\penalty0 161--168, 2016.

\bibitem[Graffy et~al.(2019)Graffy, Liu, O’Connor, Summers, and Pickhardt]{graffy2019automated}
Peter~M Graffy, Jiamin Liu, Stacy O’Connor, Ronald~M Summers, and Perry~J Pickhardt.
\newblock Automated segmentation and quantification of aortic calcification at abdominal ct: application of a deep learning-based algorithm to a longitudinal screening cohort.
\newblock \emph{Abdominal Radiology}, 44:\penalty0 2921--2928, 2019.

\bibitem[Morganti et~al.(2014)Morganti, Conti, Aiello, Valentini, Mazzola, Reali, and Auricchio]{morganti2014simulation}
Simone Morganti, Michele Conti, M~Aiello, A~Valentini, A~Mazzola, A~Reali, and F~Auricchio.
\newblock Simulation of transcatheter aortic valve implantation through patient-specific finite element analysis: two clinical cases.
\newblock \emph{Journal of biomechanics}, 47\penalty0 (11):\penalty0 2547--2555, 2014.

\bibitem[Loureiro-Ga et~al.(2020)Loureiro-Ga, Veiga, Fdez-Manin, Jimenez, Calvo-Iglesias, and Iniguez]{loureiro2020biomechanical}
Marcos Loureiro-Ga, Cesar Veiga, Generosa Fdez-Manin, Victor~Alfonso Jimenez, Francisco Calvo-Iglesias, and Andres Iniguez.
\newblock A biomechanical model of the pathological aortic valve: simulation of aortic stenosis.
\newblock \emph{Computer Methods in Biomechanics and Biomedical Engineering}, 23\penalty0 (8):\penalty0 303--311, 2020.

\bibitem[Russ et~al.(2013)Russ, Hopf, Hirsch, S{\"u}ndermann, Falk, Sz{\'e}kely, and Gessat]{russ2013simulation}
Christoph Russ, Raoul Hopf, Sven Hirsch, Simon S{\"u}ndermann, Volkmar Falk, G{\'a}bor Sz{\'e}kely, and Michael Gessat.
\newblock Simulation of transcatheter aortic valve implantation under consideration of leaflet calcification.
\newblock In \emph{2013 35th Annual International Conference of the IEEE Engineering in Medicine and Biology Society (EMBC)}, pages 711--714. IEEE, 2013.

\bibitem[Bianchi et~al.(2019)Bianchi, Marom, Ghosh, Rotman, Parikh, Gruberg, and Bluestein]{bianchi2019patient}
Matteo Bianchi, Gil Marom, Ram~P Ghosh, Oren~M Rotman, Puja Parikh, Luis Gruberg, and Danny Bluestein.
\newblock Patient-specific simulation of transcatheter aortic valve replacement: impact of deployment options on paravalvular leakage.
\newblock \emph{Biomechanics and modeling in mechanobiology}, 18:\penalty0 435--451, 2019.

\bibitem[Kong et~al.(2021)Kong, Wilson, and Shadden]{kong2021deep}
Fanwei Kong, Nathan Wilson, and Shawn Shadden.
\newblock A deep-learning approach for direct whole-heart mesh reconstruction.
\newblock \emph{Medical image analysis}, 74:\penalty0 102222, 2021.

\bibitem[Kong and Shadden(2022)]{kong2022learning}
Fanwei Kong and Shawn~C Shadden.
\newblock Learning whole heart mesh generation from patient images for computational simulations.
\newblock \emph{IEEE Transactions on Medical Imaging}, 2022.

\bibitem[Pak et~al.(2021{\natexlab{a}})Pak, Liu, Kim, Liang, McKay, Sun, and Duncan]{pak2021distortion}
Daniel~H Pak, Minliang Liu, Theodore Kim, Liang Liang, Raymond McKay, Wei Sun, and James~S Duncan.
\newblock Distortion energy for deep learning-based volumetric finite element mesh generation for aortic valves.
\newblock In \emph{Medical Image Computing and Computer Assisted Intervention--MICCAI 2021: 24th International Conference, Strasbourg, France, September 27--October 1, 2021, Proceedings, Part VI 24}, pages 485--494. Springer, 2021{\natexlab{a}}.

\bibitem[Pak et~al.(2021{\natexlab{b}})Pak, Liu, Ahn, Caballero, Onofrey, Liang, Sun, and Duncan]{pak2021weakly}
Daniel~H Pak, Minliang Liu, Shawn~S Ahn, Andr{\'e}s Caballero, John~A Onofrey, Liang Liang, Wei Sun, and James~S Duncan.
\newblock Weakly supervised deep learning for aortic valve finite element mesh generation from 3d ct images.
\newblock In \emph{Information Processing in Medical Imaging: 27th International Conference, IPMI 2021, Virtual Event, June 28--June 30, 2021, Proceedings 27}, pages 637--648. Springer, 2021{\natexlab{b}}.

\bibitem[Pak et~al.(2023)Pak, Liu, Kim, Liang, Caballero, Onofrey, Ahn, Xu, McKay, Sun, et~al.]{pak2023patient}
Daniel~H Pak, Minliang Liu, Theodore Kim, Liang Liang, Andres Caballero, John Onofrey, Shawn~S Ahn, Yilin Xu, Raymond McKay, Wei Sun, et~al.
\newblock Patient-specific heart geometry modeling for solid biomechanics using deep learning.
\newblock \emph{IEEE Transactions on Medical Imaging}, 2023.

\bibitem[Fu et~al.(2015)Fu, Liu, and Guo]{fu2015computing}
Xiao-Ming Fu, Yang Liu, and Baining Guo.
\newblock Computing locally injective mappings by advanced mips.
\newblock \emph{ACM Transactions on Graphics (TOG)}, 34\penalty0 (4):\penalty0 1--12, 2015.

\bibitem[Ronneberger et~al.(2015)Ronneberger, Fischer, and Brox]{ronneberger2015u}
Olaf Ronneberger, Philipp Fischer, and Thomas Brox.
\newblock U-net: Convolutional networks for biomedical image segmentation.
\newblock In \emph{Medical Image Computing and Computer-Assisted Intervention--MICCAI 2015: 18th International Conference, Munich, Germany, October 5-9, 2015, Proceedings, Part III 18}, pages 234--241. Springer, 2015.

\bibitem[Isensee et~al.(2021)Isensee, Jaeger, Kohl, Petersen, and Maier-Hein]{isensee2021nnu}
Fabian Isensee, Paul~F Jaeger, Simon~AA Kohl, Jens Petersen, and Klaus~H Maier-Hein.
\newblock nnu-net: a self-configuring method for deep learning-based biomedical image segmentation.
\newblock \emph{Nature methods}, 18\penalty0 (2):\penalty0 203--211, 2021.

\bibitem[Shen et~al.(2021)Shen, Gao, Yin, Liu, and Fidler]{shen2021deep}
Tianchang Shen, Jun Gao, Kangxue Yin, Ming-Yu Liu, and Sanja Fidler.
\newblock Deep marching tetrahedra: a hybrid representation for high-resolution 3d shape synthesis.
\newblock \emph{Advances in Neural Information Processing Systems}, 34:\penalty0 6087--6101, 2021.

\bibitem[Hang(2015)]{hang2015tetgen}
Si~Hang.
\newblock Tetgen, a delaunay-based quality tetrahedral mesh generator.
\newblock \emph{ACM Trans. Math. Softw}, 41\penalty0 (2):\penalty0 11, 2015.

\bibitem[Payne and Toga(1990)]{payne1990surface}
Bradley~A Payne and Arthur~W Toga.
\newblock Surface mapping brain function on 3d models.
\newblock \emph{IEEE Computer Graphics and Applications}, 10\penalty0 (5):\penalty0 33--41, 1990.

\bibitem[Chan and Purisima(1998)]{chan1998new}
Shek~Ling Chan and Enrico~O Purisima.
\newblock A new tetrahedral tesselation scheme for isosurface generation.
\newblock \emph{Computers \& Graphics}, 22\penalty0 (1):\penalty0 83--90, 1998.

\bibitem[Valette et~al.(2008)Valette, Chassery, and Prost]{valette2008generic}
S{\'e}bastien Valette, Jean~Marc Chassery, and R{\'e}my Prost.
\newblock Generic remeshing of 3d triangular meshes with metric-dependent discrete voronoi diagrams.
\newblock \emph{IEEE Transactions on Visualization and Computer Graphics}, 14\penalty0 (2):\penalty0 369--381, 2008.

\bibitem[Martin and Sun(2015)]{martin2015comparison}
Caitlin Martin and Wei Sun.
\newblock Comparison of transcatheter aortic valve and surgical bioprosthetic valve durability: a fatigue simulation study.
\newblock \emph{Journal of biomechanics}, 48\penalty0 (12):\penalty0 3026--3034, 2015.

\bibitem[Mao et~al.(2018)Mao, Wang, Kodali, and Sun]{mao2018numerical}
Wenbin Mao, Qian Wang, Susheel Kodali, and Wei Sun.
\newblock Numerical parametric study of paravalvular leak following a transcatheter aortic valve deployment into a patient-specific aortic root.
\newblock \emph{Journal of biomechanical engineering}, 140\penalty0 (10):\penalty0 101007, 2018.

\bibitem[Caballero et~al.(2020)Caballero, Mao, McKay, and Sun]{caballero2020impact}
Andr{\'e}s Caballero, Wenbin Mao, Raymond McKay, and Wei Sun.
\newblock The impact of self-expandable transcatheter aortic valve replacement on concomitant functional mitral regurgitation: a comprehensive engineering analysis.
\newblock \emph{Structural Heart}, 4\penalty0 (3):\penalty0 179--191, 2020.

\bibitem[{Medtronic LLC}(2017)]{medtronic2017corevalve}
{Medtronic LLC}.
\newblock Medtronic corevalve system instructions for use.
\newblock \url{https://www.accessdata.fda.gov/cdrh_docs/pdf13/P130021S033C.pdf}, 2017.
\newblock Accessed: 2023-09-04.

\bibitem[Scott(2015)]{scott2015multivariate}
David~W Scott.
\newblock \emph{Multivariate density estimation: theory, practice, and visualization}.
\newblock John Wiley \& Sons, 2015.

\bibitem[Virtanen et~al.(2020)Virtanen, Gommers, Oliphant, Haberland, Reddy, Cournapeau, Burovski, Peterson, Weckesser, Bright, {van der Walt}, Brett, Wilson, Millman, Mayorov, Nelson, Jones, Kern, Larson, Carey, Polat, Feng, Moore, {VanderPlas}, Laxalde, Perktold, Cimrman, Henriksen, Quintero, Harris, Archibald, Ribeiro, Pedregosa, {van Mulbregt}, and {SciPy 1.0 Contributors}]{2020SciPy-NMeth}
Pauli Virtanen, Ralf Gommers, Travis~E. Oliphant, Matt Haberland, Tyler Reddy, David Cournapeau, Evgeni Burovski, Pearu Peterson, Warren Weckesser, Jonathan Bright, St{\'e}fan~J. {van der Walt}, Matthew Brett, Joshua Wilson, K.~Jarrod Millman, Nikolay Mayorov, Andrew R.~J. Nelson, Eric Jones, Robert Kern, Eric Larson, C~J Carey, {\.I}lhan Polat, Yu~Feng, Eric~W. Moore, Jake {VanderPlas}, Denis Laxalde, Josef Perktold, Robert Cimrman, Ian Henriksen, E.~A. Quintero, Charles~R. Harris, Anne~M. Archibald, Ant{\^o}nio~H. Ribeiro, Fabian Pedregosa, Paul {van Mulbregt}, and {SciPy 1.0 Contributors}.
\newblock {{SciPy} 1.0: Fundamental Algorithms for Scientific Computing in Python}.
\newblock \emph{Nature Methods}, 17:\penalty0 261--272, 2020.
\newblock \doi{10.1038/s41592-019-0686-2}.

\bibitem[Zhuang and Shen(2016)]{zhuang2016multi}
Xiahai Zhuang and Juan Shen.
\newblock Multi-scale patch and multi-modality atlases for whole heart segmentation of mri.
\newblock \emph{Medical image analysis}, 31:\penalty0 77--87, 2016.

\bibitem[Drozdzal et~al.(2016)Drozdzal, Vorontsov, Chartrand, Kadoury, and Pal]{drozdzal2016importance}
Michal Drozdzal, Eugene Vorontsov, Gabriel Chartrand, Samuel Kadoury, and Chris Pal.
\newblock The importance of skip connections in biomedical image segmentation.
\newblock In \emph{International Workshop on Deep Learning in Medical Image Analysis, International Workshop on Large-Scale Annotation of Biomedical Data and Expert Label Synthesis}, pages 179--187. Springer, 2016.

\bibitem[Sudre et~al.(2017)Sudre, Li, Vercauteren, Ourselin, and Jorge~Cardoso]{sudre2017generalised}
Carole~H Sudre, Wenqi Li, Tom Vercauteren, Sebastien Ourselin, and M~Jorge~Cardoso.
\newblock Generalised dice overlap as a deep learning loss function for highly unbalanced segmentations.
\newblock In \emph{Deep Learning in Medical Image Analysis and Multimodal Learning for Clinical Decision Support: Third International Workshop, DLMIA 2017, and 7th International Workshop, ML-CDS 2017, Held in Conjunction with MICCAI 2017, Qu{\'e}bec City, QC, Canada, September 14, Proceedings 3}, pages 240--248. Springer, 2017.

\bibitem[Ma et~al.(2021)Ma, Chen, Ng, Huang, Li, Li, Yang, and Martel]{ma2021loss}
Jun Ma, Jianan Chen, Matthew Ng, Rui Huang, Yu~Li, Chen Li, Xiaoping Yang, and Anne~L Martel.
\newblock Loss odyssey in medical image segmentation.
\newblock \emph{Medical Image Analysis}, 71:\penalty0 102035, 2021.

\bibitem[Paszke et~al.(2017)Paszke, Gross, Chintala, Chanan, Yang, DeVito, Lin, Desmaison, Antiga, and Lerer]{paszke2017automatic}
Adam Paszke, Sam Gross, Soumith Chintala, Gregory Chanan, Edward Yang, Zachary DeVito, Zeming Lin, Alban Desmaison, Luca Antiga, and Adam Lerer.
\newblock Automatic differentiation in pytorch.
\newblock 2017.

\bibitem[Kervadec et~al.(2019)Kervadec, Bouchtiba, Desrosiers, Granger, Dolz, and Ayed]{kervadec2019boundary}
Hoel Kervadec, Jihene Bouchtiba, Christian Desrosiers, Eric Granger, Jose Dolz, and Ismail~Ben Ayed.
\newblock Boundary loss for highly unbalanced segmentation.
\newblock In \emph{International conference on medical imaging with deep learning}, pages 285--296. PMLR, 2019.

\bibitem[Wang et~al.(2018)Wang, Zhang, Li, Fu, Liu, and Jiang]{wang2018pixel2mesh}
Nanyang Wang, Yinda Zhang, Zhuwen Li, Yanwei Fu, Wei Liu, and Yu-Gang Jiang.
\newblock Pixel2mesh: Generating 3d mesh models from single rgb images.
\newblock In \emph{Proceedings of the European conference on computer vision (ECCV)}, pages 52--67, 2018.

\bibitem[Schroeder et~al.(1998)Schroeder, Martin, and Lorensen]{schroeder1998visualization}
Will Schroeder, Kenneth~M Martin, and William~E Lorensen.
\newblock \emph{The visualization toolkit an object-oriented approach to 3D graphics}.
\newblock Prentice-Hall, Inc., 1998.

\bibitem[Treece et~al.(1999)Treece, Prager, and Gee]{treece1999regularised}
Graham~M Treece, Richard~W Prager, and Andrew~H Gee.
\newblock Regularised marching tetrahedra: improved iso-surface extraction.
\newblock \emph{Computers \& Graphics}, 23\penalty0 (4):\penalty0 583--598, 1999.

\end{thebibliography}

\end{document}